\documentclass[12pt,preprint]{aastex}
\usepackage{lscape}

\newcommand{\Mth}{M_{\rm th}}
\newcommand{\rs}{r_{\rm s}}
\newcommand{\phip}{\Phi_{\rm p}}
\newcommand{\g}{{\rm G}}
\newcommand{\cs}{c_{\rm s}}
\newcommand{\rp}{r_{\rm p}}
\newcommand{\mplanet}{M_{\rm p}}
\newcommand{\lsh}{l_{\rm sh}}

\newcommand{\zp}{\zeta_{\rm peak}}
\def\ba{\begin{eqnarray}}
\def\ea{\end{eqnarray}}

\begin{document}
\title{Density Waves Excited by Low-Mass Planets in Protoplanetary Disks II: High-Resolution Simulations of the Nonlinear Regime}

\shorttitle{Nonlinear Stage of the Density Waves in protostellar Disks}

\shortauthors{Dong, Rafikov \& Stone}

\author{Ruobing Dong, Roman R. Rafikov\altaffilmark{1}, and James M. Stone}

\affil{\it Department of Astrophysical Sciences, Princeton University, Princeton, NJ 08544; rdong@astro.princeton.edu, rrr@astro.princeton.edu, jstone@astro.princeton.edu}
\altaffiltext{1}{Sloan Fellow}

\begin{abstract}

We investigate numerically the propagation of density waves excited by a low-mass
planet in a protoplanetary disk in the nonlinear regime, using 2D local shearing box
simulations with the grid-based code Athena at high spatial resolution (256 grid
points per scale height $h$). The nonlinear evolution results in the wave steepening into a shock,
causing damping and angular momentum transfer to the disk. On long timescales
this leads to spatial redistribution of the disk density, causing migration feedback
and potentially resulting in gap opening. Previous numerical studies concentrated on
exploring these secondary phenomena as probes of the nonlinear wave evolution. Here
we focus on exploring the evolution of the basic wave properties, such as its
density profile evolution, shock formation, post-shock wave behavior, and provide
comparison with analytical theory. The generation of potential vorticity at the
shock is computed analytically and is subsequently verified by simulations and 
 used to pinpoint the shock location. We confirm the theoretical relation between
the shocking length and the planet mass (including the effect of the equation of
state), and the post-shock decay of the angular momentum flux carried by the wave.
The post-shock evolution of the wave profile is explored, and we quantitatively
confirm its convergence to the theoretically expected N-wave shape. The accuracy
of various numerical algorithms used to compute the nonlinear wave evolution is also
investigated: we find that higher order spatial reconstruction and high
resolution are crucial for capturing the shock formation correctly.

\end{abstract}

\keywords{Planet-disk interactions, protostellar disks, Hydrodynamics, Methods: numerical, Planets and satellites: formation}

%%%%%%%%%%%%%%%%%%%%%%%%%%%%%%%%%%%%%%%%%

\section{INTRODUCTION}\label{sec:intro}

%%%%%%%%%%%%%%%%%%%%%%%%%%%%%%%%%%%%%%%%%

In recent years the study of disk-planet interaction in protoplanetary disks, which aims at explaining the enormous diversity of the observed exoplanetary systems \citep{pap06,pap07}, has received much attention. Specifically, the discoveries of Jovian planets and brown dwarfs residing at both extremely tight \citep{may95} and wide \citep{cha04,neu05} orbits challenge the current models of planet formation, and inspire studies of planetary migration driven by disk-planet interaction.

A planet embedded in a protoplanetary disk launches spiral density waves in the disk, which propagate away from the planet. Gravitational interaction between these nonaxisymmetric density waves and the planet exchanges angular momentum between them, driving type I planetary migration for a low mass planet, which can not open a gap in the disk. The timescale for type I migration is usually very short ($\lesssim10^5$ yr for planets more massive than a few $M_\oplus$, \citealt{tan02,bat03,dan08}) compared to the typical life time of circumstellar disks (a few $\times10^6$ yr, \citealt{har98}), which makes planetary survival a major puzzle. Several solutions to this problem have been proposed, involving magnetic fields \citep{ter03}, turbulent density fluctuations caused by magnetorotational instability (MRI) \citep{nel04}, density traps \citep{men04}, and co-orbital torques \citep{paa06,bar08,paa08,kle08}, to name a few.

However, the majority of the theoretical calculations of type I migration do not take into account the change of the disk density profile caused by the planetary density waves \citep{tan02}. In reality, density waves carry angular momentum as they travel away from the planet. Eventually, this angular momentum is deposited elsewhere in the disk, leading to the redistribution of the disk mass. For a migrating low-mass planet this density redistribution slightly enhances the disk surface density in front of the planet and reduces the density behind it, slowing down or even halting the migration \citep{hou84,war89,war97,raf02b}. This effect is called {\it migration feedback}. At higher planetary masses density redistribution by planetary torques is so severe that a gap may form near the planetary orbit. This shifts the planetary migration from type I to type II, and effectively slows the migration down, which may save planets from falling onto the central star \citep{lin86a,lin86b,war97}.

It is important to emphasize that the change of the state of the disk by planetary density waves can only be accomplished by virtue of some damping processes \citep{gol89}, which can be either linear or nonlinear. Several possible linear wave damping mechanisms have been proposed, such as radiative damping \citep{cas96} and viscous damping \citep{tak96}. However, all of these mechanisms have drawbacks and may not be very efficient at dissipating angular momentum carried by the planet-generated density waves \citep[hereafter GR01]{goo01}.

GR01 proposed another wave damping process, which results from the nonlinear wave evolution into a shock. This process could serve as a universal nonlinear wave damping mechanism working at regimes where alternatives fail (specifically, it does not require a background disk viscosity) (\citealt{lar89,lar90}, GR01). GR01 analytically investigated the nonlinear propagation of the density wave in the local shearing-sheet approximation, and predicted the dependence of the shocking length $\lsh$, the radial separation between the orbit of the planet and the point where the shock first appears, on the equation of state and the planet mass. Subsequently, \citet{raf02a} extended the local treatment of GR01 into the global case to include the effects of surface density and temperature variations in the disk as well as the disk cylindrical geometry and nonuniform shear.

Previously, \citet{lin93} have suggested the following condition for the gap opening:
\begin{equation}
\mplanet\gtrsim\Mth= \frac{\cs^3}{\Omega G}
\label{eq:mth}
\end{equation}
where $\cs$ is the sound speed in the disk and $\Omega_p$ is the angular velocity of the planet. The thermal mass $\Mth$ is the mass of a planet at which the Hill radius and the Bondi radius $R_B\equiv G\mplanet/c_s^2$ of the planet are comparable to the scale height of the gaseous disk $h$ \citep{raf06}. For a MMSN model \citep[Minimum Mass Solar Nebulae]{hay81} and $M_\star=M_\odot$
\begin{equation}
\Mth\approx 12
\left(\frac{c}{1~{\rm km~s^{-1}}}\right)^3
\left(\frac{r_{\rm p}}{1~{\rm AU}}\right)^{3/4}~
 M_\oplus,
\label{eq:m1}
\end{equation}
where $r_{\rm p}$ is the semi-major axis of the planetary orbit.

GR01 found that planets with $\mplanet\gtrsim\Mth$ generate density waves which are nonlinear from the very start, and shock as soon as they are produced. Their angular momentum is deposited in the immediate vicinity of the planet, within $(1\sim2)h$, leading to the surface density evolution and gap opening close to the planet. On the other hand, \citet{raf02b} demonstrated that density waves produced by lower mass planets are still able to dissipate efficiently further out through the nonlinear damping, even if they are only weakly nonlinear to begin with. In fact one should expect that given enough time, even a very low mass planet at a fixed semi-major axis should be able to open a gap in an inviscid disk. Based on this argument, \citet{raf02b} explored gap opening mediated by the nonlinear dissipation of density waves and indeed found that the low mass planets which do not satisfy the condition~(\ref{eq:mth}) can be capable of stalling their migration via the migration feedback and opening a gap in a low viscosity disk.

However, most previous numerical studies have failed to capture gap opening by small planets ($\mplanet\lesssim\Mth$). This is most likely because these simulations have studied flows with significant viscosity (artificial or numerical). In a viscous disk there is another condition \citep{lin93,raf02b}
\begin{equation}
M_p\gtrsim\Mth\left(\frac{\alpha}{0.043}\frac{a}{h}\right)^{1/2}
\label{eq:mvis}
\end{equation}
that has to be fulfilled simultaneously with (\ref{eq:mth}) for the gap to open, where the effective Shakura-Sunyaev $\alpha=\nu/h\cs$ ($\nu$ is the kinematic viscosity). It is quite likely that the levels of viscosity in these simulations were too high (in particular, the numerical viscosity due to the low spatial resolution) for the torques induced by the low mass planets to overcome viscous filling of the gap, {\it i.e.} the viscous criterion (\ref{eq:mvis}) was not fulfilled.

Only recently has the GR01 theory been confirmed in numerical work. \citet{paar} first found evidence for the distance over which density waves damp to {\it increase} with decreasing $\mplanet$, exactly as predicted by GR01. Subsequently, using two-dimensional hydrodynamic simulations of migration of low-mass planets in nearly inviscid disks, \citet{li09} found the migration rate to drop due to the migration feedback. They also showed that high disk viscosity ($\alpha\gtrsim10^{-3}$) could wash out this effect, making fast type I migration persist. In addition, they confirmed the existence of a critical planet mass which eventually halts the migration, and found reasonable agreement with the theoretical prediction of \citet{raf02b}. \citet{mut10} also investigated nonlinear wave evolution by 2D hydrodynamical simulations and confirmed that low mass planets (a few tenths of $\Mth$) indeed are able to open gaps in an inviscid disk. They successfully detected shock formation and density redistribution in the disk, resulting from the nonlinear wave evolution. However, these simulations investigated only {\it secondary} consequences of the nonlinear wave evolution, such as the slowdown of the migration due to migration feedback and the limit on gap opening planet mass.

It is obvious from this discussion that elucidating the mechanisms of nonlinear wave dissipation in protoplanetary disks is not only interesting in and of itself, but is also crucial for understanding the migration feedback and the gap opening issue, both of which could significantly increase the planetary migration timescale. Here, following \citet[here-after Paper I]{don11}, who explored the linear excitation and propagation of planet-generated density waves, we continue to investigate the evolution of density waves via numerical simulations. We now focus on the nonlinear wave evolution, and provide detailed quantitative comparisons with analytical results of GR01.

The structure of this paper is as follows. In \S~\ref{sec:theory}, we briefly summarize the main results of the nonlinear theory, and analytically study the potential vorticity generation at the shock, which we use to pinpoint the shock location. The code and numerical setup of the simulations are introduced in \S~\ref{sec:code}. We then present the main numerical results and compare them with theory in \S~\ref{sec:result}. The effects of variation of numerical algorithms are investigated in \S~\ref{sec:variety}, and the effect of linear damping due to viscosity is studied in \S~\ref{sec:viscosity}. We discuss the implications for realistic protoplanetary disks in \S~\ref{sec:implications}, and summarize main results in \S~\ref{sec:summary}.

%%%%%%%%%%%%%%%%%%%%%%%%%%%%%%%%%%%%%%%%%

\section{Nonlinear Theory of the Density Wave Propagation}
\label{sec:theory}

%%%%%%%%%%%%%%%%%%%%%%%%%%%%%%%%%%%%%%%%%

Here we briefly summarize the results of GR01 for the nonlinear density wave evolution. For a planet of sufficiently low mass ({\it i.e.} $\mplanet\ll\Mth$), the excitation and initial propagation of the wave are linear processes, which are not affected by the nonlinearity. Far from the planet, the wave excitation is no longer important, while the nonlinear effects start to accumulate. In the shearing sheet geometry (as usual, $x=r-r_{\rm p}$ and $y=r_{\rm p}(\theta-\theta_{\rm p})$ denote pseudo-Cartesian radial and azimuthal coordinates in a corotating system centered on the planet, and {\bf $(u,v)$} represents the perturbed velocity in $(x,y)$ plane) GR01 have shown that under the assumption of weak nonlinearity the fully nonlinear system of fluid equation can be reduced to a single first-order nonlinear equation:
\begin{equation}
\label{eq:chieqn}
\partial_{t}\chi-\chi\partial_\eta\chi = 0,
\end{equation}
which is the inviscid Burger's equation \citep{whi74}. The dimensionless variables appearing here are related to radius, azimuth, and density contrast as follows:
\begin{eqnarray}
{t} &\equiv&  
%\frac{6}{5}\left|\frac{y}{2h/3}\right|^{5/4}\frac{\mplanet}{\Mth}
\frac{3^{7/2}}{2^{11/4}5}\left|\frac{x}{h}\right|^{5/2}\frac{\mplanet}{\Mth}, 
\label{eq:t} \\
\eta &\equiv& \frac{3}{2}\left[\frac{y}{h}+
\frac{3x^2}{4h^2}\mbox{sign}(x)\right], 
\label{eq:eta} \\
\chi &\equiv& 
%(\gamma+1)\left|\frac{y}{2h/3}\right|^{-1/4}
%\frac{\Sigma-\Sigma_0}{\Sigma_0}\frac{\Mth}{3\mplanet}~.
\frac{2^{3/4}(\gamma+1)}{3^{3/2}}\left|\frac{h}{x}\right|^{1/2}
\frac{\Sigma-\Sigma_0}{\Sigma_0}\frac{\Mth}{\mplanet}~.
\label{eq:chi} 
\end{eqnarray}
where $h$ is the scale height and $\cs$ is the sound speed of the disk, 
and $\gamma$ is the adiabatic index in the equation of state.
\citet{paar} has demonstrated that in the case of an isothermal disk in definition
(\ref{eq:chi}) $(\Sigma-\Sigma_0)/\Sigma_0$ should be replaced with $\ln(\Sigma/\Sigma_0)$.
However, since we are interested only in the {\it weakly} nonlinear wave
evolution the assumption of $(\Sigma-\Sigma_0)/\Sigma_0\ll 1$ is always 
made and Eq. (\ref{eq:chi}) holds even in the isothermal case.

In the absence of linear damping mechanisms such as viscosity, for almost any choice of smooth initial conditions in Burger's Eq. (\ref{eq:chieqn}), the solution will eventually become double-valued, which means that a shock must appear. This phenomenon is analogous to the nonlinear evolution of sound waves, except that for planetary density waves (in the linear regime) the conservation of the wave angular momentum flux (AMF) and differential rotation lead to the wave amplitude {\it increasing}, and the radial wavelength {\it decreasing} with distance from the excitation point in the linear regime. Both effects conspire to accelerate wave steepening, and the formation of shocks close to the planet. GR01 provided an analytical formula for the nonlinear shocking length $\lsh$:
\begin{equation}
\lsh\approx0.8 \left(\frac{\gamma+1}{12/5}\frac{ \mplanet}{\Mth}\right)^{-2/5}h,
\label{eq:ls-mp}
\end{equation}
which indicates that if $\mplanet\gtrsim\Mth$, the shock appears very close to planet ($\lsh\sim h$), {\it i.e.} in the region where the wave excitation is taking place. GR01 also predicted that deep into the post shock regime ($x\gg\lsh$) the density wave profile should attain the N-wave shape \citep{lan59}. The amplitude of the N-wave scaled by $x^{1/2}$
\begin{equation}
\Delta=\frac{\delta\Sigma}{\Sigma_0}\left|\frac{h}{x}\right|^{1/2}
\frac{\Mth}{\mplanet}
\label{eq:delta}
\end{equation}
and its azimuthal width $w$ should scale with distance from the planet as:
\begin{equation}
\Delta\propto t^{-1/2},\ \ \ \ w\propto t^{1/2},
\label{eq:nonlinearwave}
\end{equation}
where the time-like coordinate $t$ is defined by Eq. (\ref{eq:t}). Note that $\Delta\propto t^0$ (stays roughly constant) in the linear regime of wave evolution due to the angular momentum conservation.

After the shock formation, the wave gradually damps out, transferring its angular momentum to the mean flow, and forcing the disk to evolve. The angular momentum flux $F_H(x)$ carried by the wave decays in the post-shock region as (GR01):
\begin{equation}
F_H(x)\propto |x|^{-5/4}\quad(|x|\gg \lsh)
\label{eq:amf-decay}
\end{equation}
Where the shearing-sheet geometry is assumed. In the global setting the post-shock wave evolution was investigated 
in \citet{raf02a}.

%%%%%%%%%%%%%%%%%%%%%%%%%%%%%%%%%%%%%%%%%%% 

\subsection{Generation of potential vorticity}
\label{sec:pvtheory}

A very useful diagnostic of the
density wave evolution is provided by the so-called {\it potential vorticity}, 
sometimes also called {\it vortensity}. This quantity is defined
in the shearing box coordinates as
\ba
\zeta\equiv\frac{{\bf e}_z\cdot\left(\nabla\times {\bf v}\right)+
2\Omega}{\Sigma},
\label{eq:pv}
\ea
where ${\bf v}$ is the fluid velocity in the frame of the shearing
sheet, and ${\bf e}_z$ is the unit vector perpendicular to the disk 
plane. The background value of this quantity for ${\bf v}=0$ is 
$\zeta_0=\Omega/(2\Sigma_0)$. In an inviscid, barotropic fluid, potential 
vorticity is conserved everywhere, except across
shocks. At the shock $\zeta$ experiences a
jump $\Delta\zeta$ the magnitude of which depends on the strength of the
shock and the orientation of the shock front with respect to
the incoming flow. Thus, the potential vorticity perturbation 
$\Delta\zeta=\zeta-\zeta_0$ can be interpreted as the evidence for the
appearance of shock waves and provides a useful diagnostic of the 
flow. 

Here we theoretically calculate the behavior of $\Delta\zeta$ as 
a function of distance from the planet and planetary mass $M_p$
based on the weakly-nonlinear theory of GR01. We start with the
following expression for the jump in $\zeta$ at the isothermal 
shock \citep{kevlahan,li05,lin10}:
\ba
\Delta\zeta=\frac{c_s}{\Sigma}\frac{\left(M^2-1\right)^2}{M^4}
\frac{\partial M}{\partial S}=
\frac{c_s}{2\Sigma M^5}\left(\frac{\Delta\Sigma}{\Sigma}\right)^2
\frac{\partial }{\partial S}\left(\frac{\Delta\Sigma}{\Sigma}\right),
\label{eq:gen_expr}
\ea
where $M$ is the Mach number of the flow perpendicular to the 
shock, $\Delta\Sigma$ is the surface density jump at the 
shock front and $S$ is the distance along the shock. To arrive 
at the last expression we used the relation 
$M^2-1=\Delta\Sigma/\Sigma$ valid in the isothermal case. 

We can write $\partial/\partial S=\left[1+(dy_{sh}/dx)^2\right]^{-1/2}
\partial/\partial x$, where $y_{sh}(x)\approx -(3/4)h(x/h)^2$ is the 
theoretical shock position in the $x-y$ plane. For $x\gtrsim h$
we find then
\ba
\frac{\partial}{\partial S}\approx \frac{2}{3}\frac{h}{x}
\frac{\partial}{\partial x}.
\label{eq:dS}
\ea
We now use Eq. (\ref{eq:chi}) to relate $\Delta\Sigma/\Sigma$
to the jump across the shock  $\Delta\chi_{sh}$ of the function $\chi$,
satisfying the Eq. (\ref{eq:chieqn}):
\ba
\frac{\Delta\Sigma}{\Sigma}=\frac{3^{3/2}}{2^{3/4}(\gamma+1)}
\left|\frac{x}{h}\right|^{1/2}
\frac{\mplanet}{\Mth}\Delta\chi_{sh}\left(t(x)\right),
\label{eq:rel}
\ea
where $t(x)$ is given by Eq. (\ref{eq:t}).

Combining 
Eqs. (\ref{eq:t}), (\ref{eq:gen_expr})-(\ref{eq:rel}),
and setting $M\approx1$ (shock is weak) in Eq. (\ref{eq:gen_expr}), we find 
\ba
\Delta\zeta(x)=\frac{3^{7/2}}{2^{13/4}(\gamma+1)^3}
\left[5\frac{\partial\ln\Delta\chi_{sh}}{\partial\ln t}+1\right]
\left[\Delta\chi_{sh}(t(x))\right]^3\frac{\Omega}{\Sigma_0}
\left|\frac{h}{x}\right|^{1/2}
\left(\frac{\mplanet}{\Mth}\right)^3.
\label{eq:dksi}
\ea
Note $\Delta\zeta\neq 0$ only for $x>l_{sh}$, where the 
shock exists. Eq. (\ref{eq:dksi}) shows that potential vorticity
generation is a steep function of $M_p$: at a fixed distance $x$ away 
from the planetary orbit $\Delta\zeta\propto\mplanet^3$. Thus, low mass 
planets are very inefficient at generating potential vorticity and
detecting its production in numerical experiments with small planets 
is a non-trivial task. 

Using Eq. (\ref{eq:ls-mp}) and setting $\gamma=1$ we can rewrite
Eq. (\ref{eq:dksi}) as:
\ba
\Delta\zeta(x/l_{sh})\approx 1.3\frac{\Omega}{\Sigma_0}
\left|\frac{l_{sh}}{x}\right|^{1/2}
\left(\frac{\mplanet}{\Mth}\right)^{16/5}
\left[\frac{\partial\ln\Delta\chi_{sh}}
{\partial\ln (x/l_{sh})}+\frac{1}{2}\right]
\left[\Delta\chi_{sh}(t(x/l_{sh}))\right]^3.
\label{eq:dksi1}
\ea
This particular form is natural since $\Delta\chi_{sh}$ is a 
function of $t\approx |x/l_{sh}|^{5/2}$ (for isothermal gas) 
only, as follows from Eq. (\ref{eq:chieqn}), which does 
not contain any free parameters. Thus, a fixed $x/l_{sh}$
corresponds to a unique value of $t$ and $\Delta\chi_{sh}$.

For instance, if we choose $x/l_{sh}$ (and $t$) to correspond 
to the point where $\Delta\chi_{sh}$ (and $\Delta\zeta$) attains 
its maximum value, the corresponding $\Delta\zeta_{max}$ would 
scale with $M_p$ as $\propto \left(\mplanet/\Mth\right)^{16/5}$.
More generally, one expects 
$\Delta\zeta\left(\Mth/\mplanet\right)^{16/5}$ to depend on 
$x/l_{sh}$ {\it only}. 

We compare these theoretical predictions with numerical
results in \S \ref{sec:pvnumerical}.

%%%%%%%%%%%%%%%%%%%%%%%%%%%%%%%%%%%%%%%%%%% 

\section{Code, Method and the Numerical setup}\label{sec:code}

%%%%%%%%%%%%%%%%%%%%%%%%%%%%%%%%%%%%%%%%%

To provide quantitative comparison with analytical theory in GR01 and in \S \ref{sec:pvtheory}, we carry out 2D shearing box hydrodynamics simulations using Athena, a grid-based code for astrophysical gas dynamics using higher-order Godunov methods \citep{gar05,gar08,sto08}. The basic numerical setup is the same as in paper I, and we briefly summarize it below.

The implementation of the shearing box approximation in Athena is described in \citet{sto10}. This approximation adopts a frame of reference located at radius $\rp$ corotating with the disk at orbital frequency $\Omega_p=\Omega(\rp)$, and the Cartesian coordinate system $(x,y)$ is the same as in \S~\ref{sec:theory}. The rate of shear corresponds to a Keplerian rotation profile. The equation of state (EOS) we use in the simulations is the ideal gas law with
\begin{equation}
  E_{\rm in} = \frac{p}{\gamma -1}
\label{eq:ein}
\end{equation}
where $E_{\rm in}$ is the internal energy, $p$ is the gas pressure, and $\gamma=5/3$. We also use an isothermal EOS (effectively $\gamma=1$), in which case $p=\Sigma\cs^2$, where $\cs$ is the isothermal sound speed. We do not include explicit viscosity except in \S~\ref{sec:viscosity} ({\it i.e.} no linear dissipation is present in the system), and we do not account for the self-gravity of the disk.

To study the accuracy and convergence of our results we have varied the numerical algorithms used to compute our simulations (\S \ref{sec:variety}). Unless noted otherwise, our standard simulations use the following choices (the same as in paper I): an isothermal equation of state, the Roe solver with third order reconstruction in characteristic variables, and the corner transport upwind (CTU) unsplit integrator \citep{sto08}, a resolution of 256 grid points per scale height $h$ (subsequently denoted $256/h$ for brevity), and fourth order accurate potential $\phip^{(4)}$
\begin{equation}
\phip^{(4)}=-\g \mplanet\frac{\rho^2+1.5\rs^2}{(\rho^2+\rs^2)^{3/2}}
\label{eq:phi4}
\end{equation}
with softening length $\rs=h/32$, where $\rho=\sqrt{x^2+y^2}$ is the cylindrical radius counted from the position of the planet. This potential converges to Keplerian at $\rho\gg \rs$ as $\left(\rs/\rho\right)^4$ (which means the fractional error is O$((r_s/\rho)^4)$ as $r_s/\rho\rightarrow0$). 

We use the following boundary conditions (BCs, the same as paper I). On $x$ (radial) boundaries, we keep values of all physical variables in ghost zones fixed at their respective unperturbed Keplerian values ({\it i.e.} keep the ghost zones as their initial states), and the waves leave through the $x$ boundaries when they reach the edge. On the $y$ (azimuthal) boundary, we experimented with two BCs: the conventional outflow BC, and an inflow/outflow BC. In the former case the variables in the ghost zones are copied from the last actively-updated row of cells. In the latter case, the variables in the ghost zones are fixed at their initial values if they are the physical ``inflow'' boundaries (the regimes $y<0, x<0$ and $y>0, x>0$), or copied from the last actively-updated row of cells if they are the physical ``outflow'' boundaries (the regimes $y>0, x<0$ and $y<0, x>0$). We found that the conventional outflow $y$ BC accumulates some non-zero velocity perturbation on top of the pure linear shear velocity profile, while our inflow/outflow $y$ BC does not. This affects calculation of variables derived from the simulated velocity field, such as potential vorticity. We use the inflow/outflow $y$ BC for our simulations.

As in paper I, we check the level of numerical viscosity in our runs. We run a series of test simulations with otherwise identical conditions but different explicit Navier-Stokes viscosity, and measure the wave properties. As the explicit viscosity decreases the simulation results gradually converge to the one with zero explicit viscosity, which indicates the numerical viscosity dominates the explicit viscosity. For a typical simulation with low $\mplanet=2.09\times 10^{-2}\Mth$ and an isothermal equation of state, the effective Shakura-Sunyaev $\alpha$-parameter ($\alpha=\nu/h\cs$) characterizing our numerical viscosity is found to be below $10^{-5}$. Such small levels of viscosity are expected in dead zones of protoplanetary disks \citep{gam96}, where magnetorotational instability (MRI) may not operate effectively \citep{flem03}. Also note even then the MRI operates, MHD turbulence may not act like a Navier-Stokes viscosity.

For the standard simulations in this work we use box size $16h\times102h$, thus the overall grid resolution in our runs is $4096\times26112$ (for the standard resolution of $256/h$). We note that these domains are larger than in paper I since we want to follow the nonlinear evolution of fluid variables deep into the post-shock regime. In a few cases with very small $\mplanet$, we extend the simulation box size to $20h\times156h$ to fit the large $\lsh$ inside the box (see Eq. \ref{eq:ls-mp}). Our simulations are run for at least 10 and in some cases up to 100 orbital periods.

%%%%%%%%%%%%%%%%%%%%%%%%%%%%%%%%%%%%%%%%%

\section{Results and Discussion}\label{sec:result}

%%%%%%%%%%%%%%%%%%%%%%%%%%%%%%%%%%%%%%%%%

In this section, we present our numerical results on the density wave properties in the nonlinear regime, and make detailed quantitative comparison with the theory.

%%%%%%%%%%%%%%%%%%%%%%%%%%%%%%%%%%%%%%%%%

\subsection{Density Profile in the Post Shock Region}
\label{sec:density}

\begin{figure}[tb]
%\vspace*{-0.5cm}
\begin{center}
\epsscale{0.51} \plotone{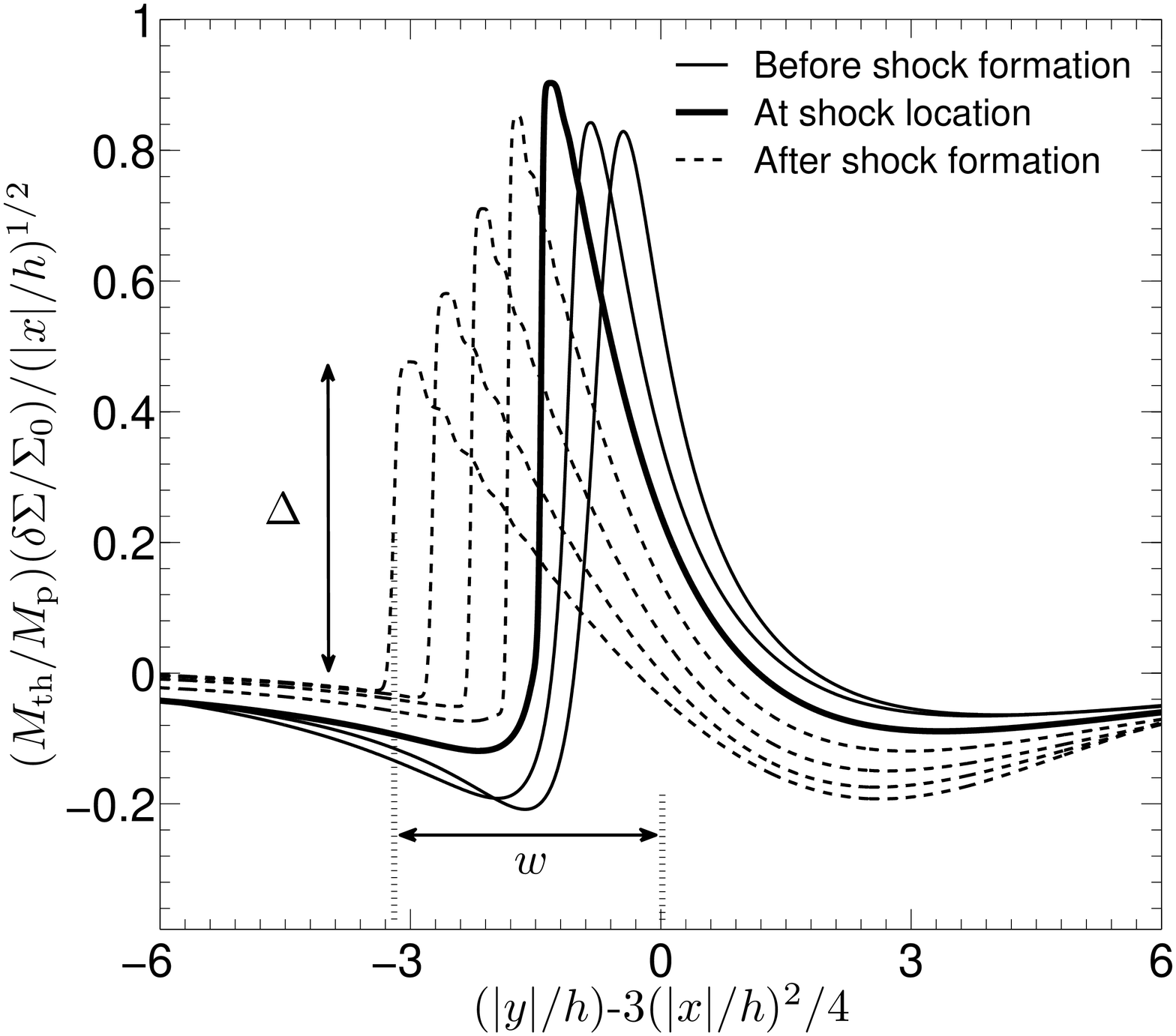} \epsscale{0.35} \plotone{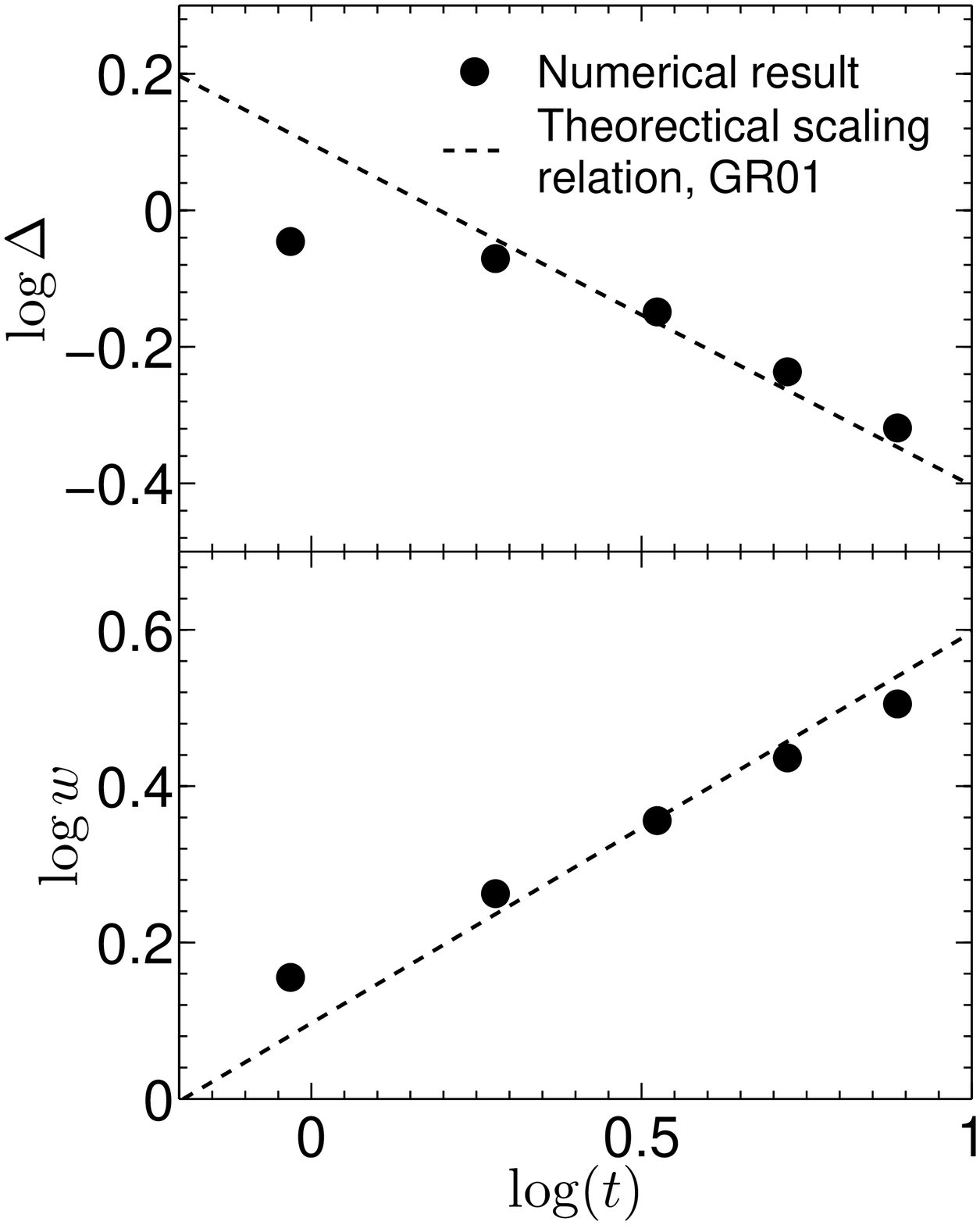}
%\vspace*{-0.5cm}
\end{center}
\figcaption{Left: Density profiles across the density wake for the $\mplanet=2.09\times10^{-2}\Mth$ (corresponding to $\lsh\approx4h$) case at two pre-shock locations ($1.33h$ and $2.67h$, thin solid curves), four post-shock locations ($5.33h$, $6.67h$, $8.0h$, and $9.33h$, dashed curves), and the theoretically predicted shocking length ($4h$, the thick solid curve). Right: Scaling of $\Delta$ and $w$ with coordinate $t(x)$ defined by Eq. (\ref{eq:t}) in the post shock region. Theoretical scaling relations (Eq. \ref{eq:nonlinearwave}) are over-plotted with arbitrary normalization.
\label{fig:density-decay}}
\end{figure}

The nonlinear evolution of the wave is depicted in Figure~\ref{fig:density-decay}a, where we show the azimuthal density profile at two pre-shock locations, the theoretically predicted shocking distance $\lsh$, and four post-shock positions. In accord with linear theory, before the shock the numerical peak amplitude of the wave stays roughly constant (we see only a slight increase with $x$ expected as a result of the continuing accumulation of the torque from low azimuthal wavenumber harmonics by the wave). At the same time the leading edge of the density profile gradually steepens as the wave propagates away from the planet. At $x\approx\lsh$ the leading edge becomes vertical and the wave shocks. After that the leading edge stays vertical while the height of the profile $\Delta$ (see Fig. \ref{fig:density-decay}a for an illustration of its definition) decreases due to the dissipation of energy and angular momentum.

We note that the N-wave shape does not quite appear in our wave profile, because the second (trailing) shock does not emerge until very large distance ($x\sim7(\Mth/\mplanet)^{2/5}h$), which our simulation boxes do not cover. Nevertheless, the leading segment of the N-wave does appear and we use its behavior to quantitatively explore the N-wave evolution. In Figure~\ref{fig:density-decay}b, we plot the dependence of $\Delta$ and the azimuthal width of the wave $w$ (the $y/h-3(x/h)^2/4$ value at the mid point of the leading edge, as indicated in Fig. \ref{fig:density-decay}a) on the coordinate $t(x)$ defined by Eq. (\ref{eq:t}) in the post shock regime, as well as the theoretical scaling relations (Eq. \ref{eq:nonlinearwave}) with arbitrary normalization. Far in the post shock regime ($t\gg1$) our numerical results on the N-wave behavior agree well with theory.

Evolution of the wave profile in the nonlinear regime was first studied by \citet{mut10}, who demonstrated the steepening of the wave profile and the decay of $\Sigma$ after the shock formation (their Figure 6). However, they did not make quantitative comparison between the numerical results and theoretical scaling relations, as we do here.

%%%%%%%%%%%%%%%%%%%%%%%%%%%%%%%%%%%%%%%%%

\subsection{Excitation of potential vorticity}
\label{sec:pvnumerical}

\begin{figure}[tb]
%\vspace*{-0.5cm}
\begin{center}
\epsscale{0.45} \plotone{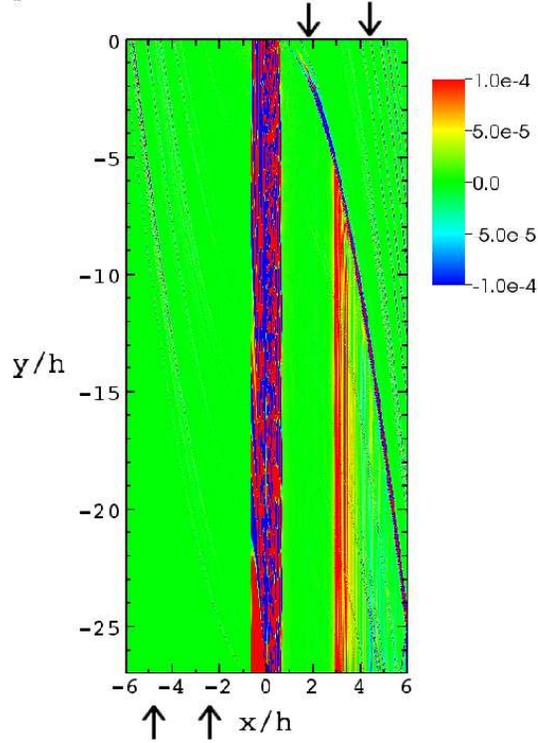}
%\vspace*{-0.5cm}
\end{center}
\figcaption{A typical snapshot of the potential vorticity perturbation $\Delta\zeta$ (normalized by $\Omega/\Sigma_0$) in shearing sheet coordinates in our simulation (only half of the simulation domain is shown). A planet with $\mplanet=4.28\times10^{-2}\Mth$ (corresponding to $\lsh\approx3h$) is located at $x=y=0$. Fluid enters from the upper right and the lower left boundaries. Note the vorticity generation at the shock position, with the amplitude of $\Delta\zeta$ decaying far from the planet (in $x$). Vorticity perturbation in the horseshoe region is not related to the shock. Narrow transient features in the preshock region are due to fluid activity in the horseshoe region.
\label{fig:pvimage}}
\end{figure}

In Figure~\ref{fig:pvimage} we show a typical spatial patten of potential vorticity deviation $\Delta\zeta=\zeta-(\Omega/2\Sigma_0$) from its background value in our simulations (only one half of the simulation box is shown). Because of the velocity shear, the fluid enters the box from the two inflow $y$ boundaries ($x>0\ \&\ y>0$ and $x<0\ \&\ y<0$) with initial background $\Delta\zeta=0$, and maintains this value until it reaches the spiral wave. If the fluid element meets a spiral wave at $|x_{\rm cross}| \lesssim\lsh$, the shock is not crossed and the original $\Delta\zeta$ value is conserved. The fluid element carries it until it leaves the simulation domain through the outflow $y$ boundary ($x<0\ \&\ y>0$ and $x<0\ \&\ y>0$).

On the other hand, if $|x_{\rm cross}|\gtrsim\lsh$, the fluid element crosses the shock, potential vorticity conservation is broken and $\zeta$ gets a kick. The amplitude of the $\zeta$ jump depends on $|x_{\rm cross}|$. A larger $|x_{\rm cross}|$ results in a weaker $\zeta$ jump because further from the planet the shock becomes weaker and the incidence angle of the fluid on the shock is more oblique. After the shock, the fluid element conserves the new value of $\zeta$ until it leaves the box. Note that $\zeta$ is ill defined right at the shock front, which produces the feature along the spiral (also visible in \citealp[Fig.1]{lin10}), and the noisy structure at small $|x|$ is due to the vortex generation in the co-orbital region \citep{kol03,lin10}. Measured as a function of $|x|$ at large $|y|\ll h$, $\Delta\zeta$ maintains the background value as small $|x|$ until $|x|\approx\lsh$, where it first grows with $x$, and then gradually decreases, as the wave decays and the shock becomes weaker and weaker.

\begin{figure}[tb]
%\vspace*{-0.5cm}
\begin{center}
\epsscale{0.45} \plotone{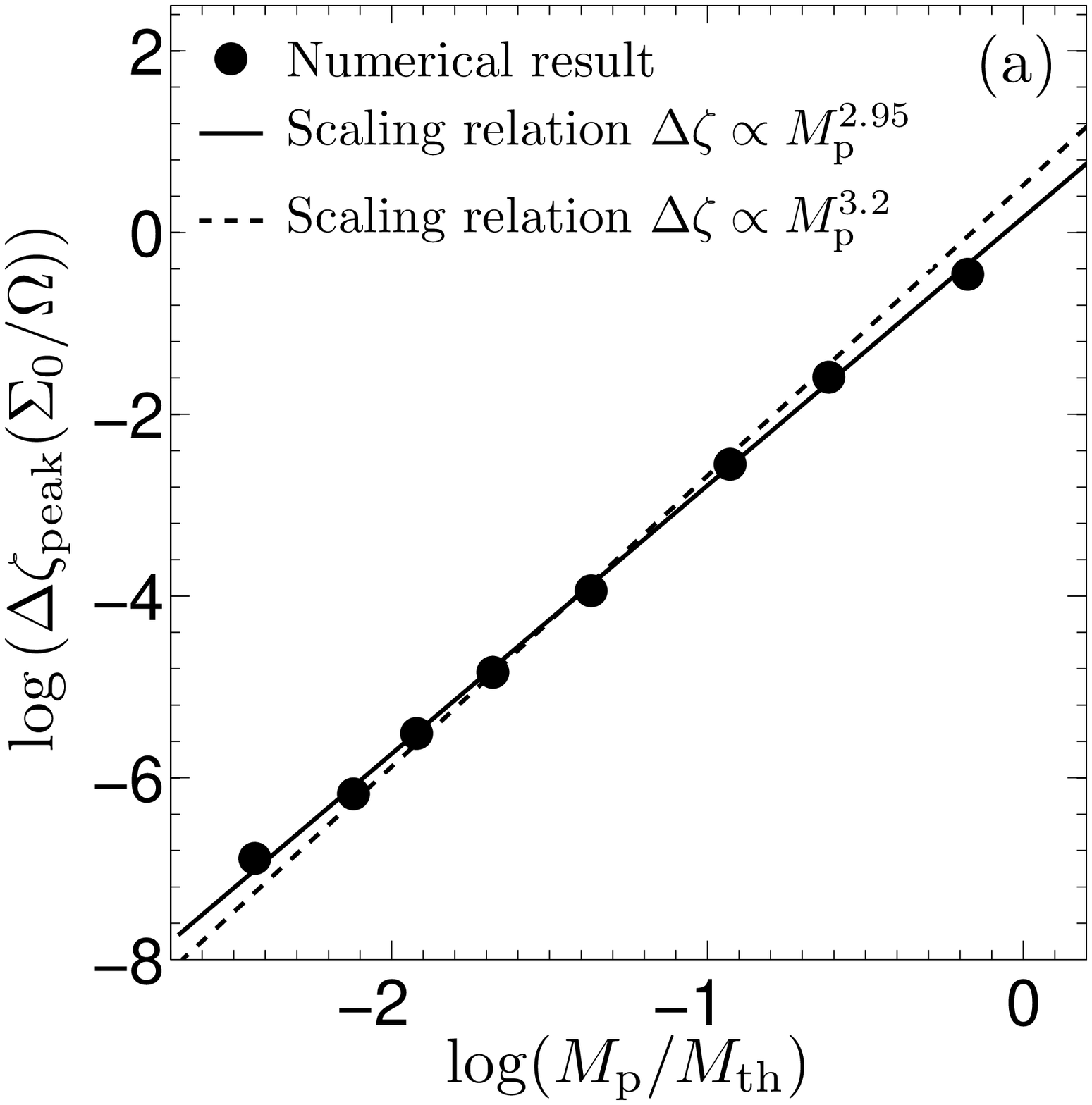} \plotone{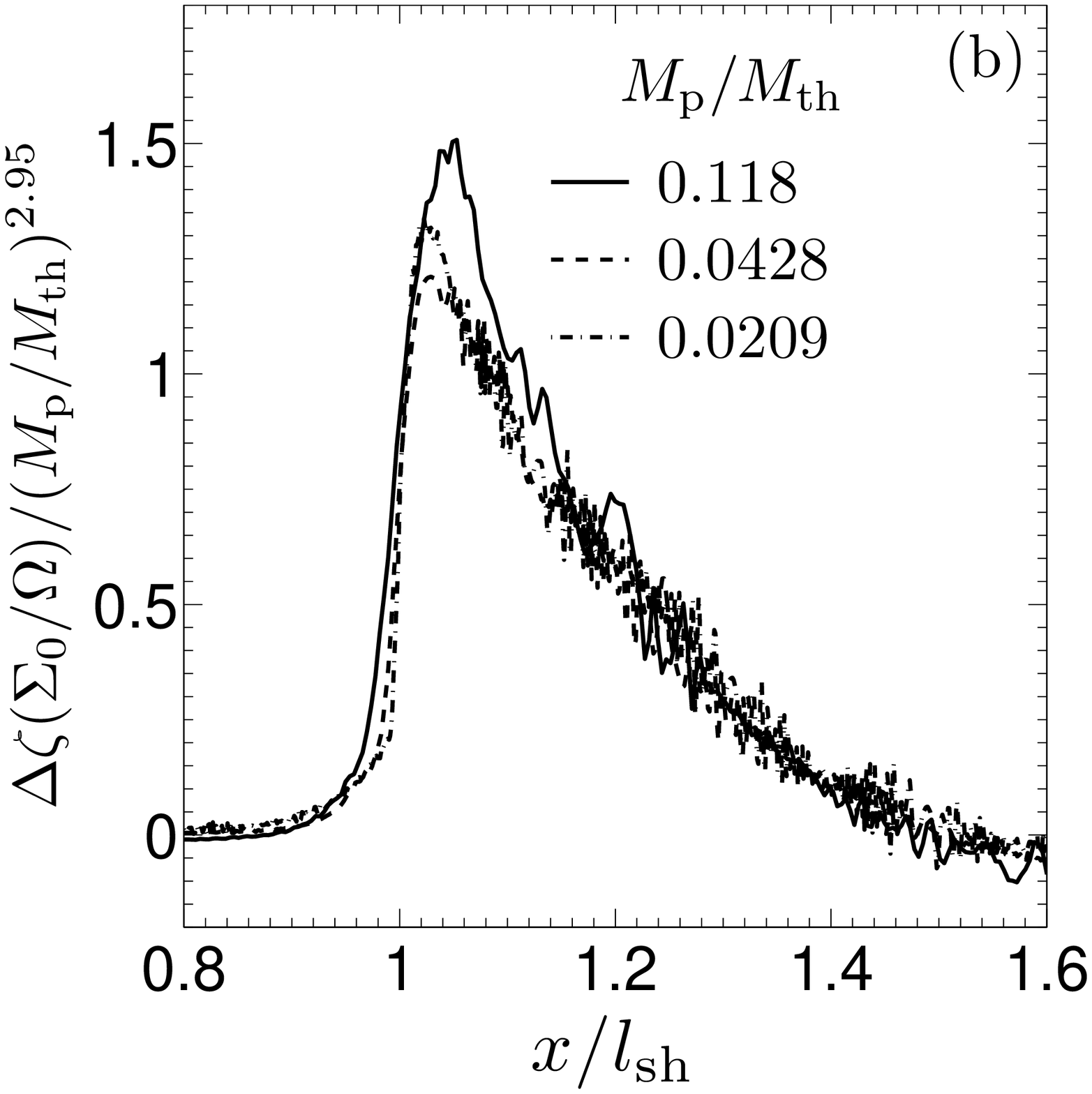}
%\vspace*{-0.5cm}
\end{center}
\figcaption{Panel (a): Numerical peak $\Delta\zp$ amplitude as a function of $\mplanet$. Best fit relation $\zeta\propto\mplanet^{2.95}$ and the theoretical scaling relation $\Delta\zp\propto\mplanet^{3.2}$ (Eq. \ref{eq:dksi1}) are shown. Panel (b): Radial profile of $\Delta\zeta$ (scaled by $\Omega/\Sigma_0$) as a function of $x$ (scaled by $\lsh$). Different curves for different $\mplanet$ have been scaled by $(\mplanet/\Mth)^{2.95}$ (the numerically best fit scaling factor) to remove the dependence on $\mplanet$. See the discussion in \S \ref{sec:pvnumerical}.
\label{fig:pvjump-mp}}
\end{figure}

In Figure~\ref{fig:pvjump-mp}a we plot the maximum value of the vorticity jump $\Delta\zeta_{max}$ (scaled by $\Omega/\Sigma_0$ to make it dimensionless) as a function of the planetary mass $\mplanet$. The value of $\Delta\zeta_{max}$ was derived by varying $x$ at fixed $y\gg y_{\rm sh}(\lsh)$ behind the shock and finding the maximum of $\Delta\zeta(x)$ at the shock in a simulation with a given $\mplanet$. Analytical calculation of $\Delta\zeta$ generation in \S \ref{sec:pvtheory} based on weakly nonlinear theory of GR01 predicts that  $\Delta\zeta_{max}\propto \mplanet^{16/5}$. However we find that our results are fit marginally better by the dependence $\Delta\zeta_{max}\propto \mplanet^{2.95}$ over more than 2 orders of magnitude in planetary mass. This result confirms the low efficiency of the potential vorticity excitation by low-mass planets.

In Figure~\ref{fig:pvjump-mp}b we plot the full radial profiles of $\Delta\zeta$ produced after a single shock crossing for three different $\mplanet$. For better graphic representation, we smooth the $\zeta$ curve by averaging value over $h/32$ in $x$ to reduce the noise. We display the data scaled by $\mplanet^{2.95}$ as a function of $x/\lsh$. The exponent $2.95$ of the mass scaling is the same as in Figure \ref{fig:pvjump-mp}a so that the resultant curve has a universal shape, independent of $\mplanet$. 

One can see that, as expected, $\Delta\zeta$ stays close to zero for $x<l_{sh}$, i.e. prior to the appearance of the shock. At the shock $\Delta\zeta$ very rapidly (within radial distance $\lesssim 0.1\l_{sh}$) attains its maximum value, as the shock front develops. Subsequently $\Delta\zeta$ decreases because the shock rapidly weakens with increasing $x$: its amplitude is reduced by the dissipation and also the shock becomes more oblique to an incoming flow. As stated in \S \ref{sec:theory} far from the planet shock evolves into an N-wave with $\chi$ decaying\footnote{In the N-wave regime one can identify $\Delta\chi_{sh}$ from Eq. (\ref{eq:rel}) with $\Delta$ defined by Eq. (\ref{eq:delta}).} as $\Delta\chi_{sh}\propto t^{-1/2}\propto (x/\lsh)^{-5/4}$, see Eqs. (\ref{eq:t}) and (\ref{eq:nonlinearwave}). As a result, according to Eq. (\ref{eq:dksi1}), at $x\gg\lsh$ the amplitude of $\Delta\zeta$ should generally decrease $\propto |x/\lsh|^{-17/4}$. 

At smaller $x\gtrsim\lsh$  $\Delta\chi_{sh}$ does not obey the N-wave scaling and decays with $x$ rather slowly, as can be directly seen from Figure \ref{fig:density-decay}b, where $\Delta$ acts as a proxy for  $\Delta\chi_{sh}$. Also, at these separations the factor in square brackets in Eq. (\ref{eq:dksi}) varies in a non-trivial fashion. Initially $\partial\ln\Delta\chi_{sh}/\partial\ln t$ is small, see Figure \ref{fig:density-decay}b, and the factor in square brackets (as well as the $\Delta\zeta$) is positive. However, at large separations N-wave evolution results in $\partial\ln\Delta\chi_{sh}/\partial\ln t\to -1/2$ and the factor in square brackets approaches $-3/2$, making  $\Delta\zeta$ negative. From our numerical calculations we find that $\Delta\zeta$ changes sign at $x\approx 1.5\lsh$. This result agrees well with numerical calculations of other authors \citep{li05}.

%\rdnote{comparison with theory. Correlation coefficient is 0.999 (for 2.95, exclude the largest $\mplanet$), so very linearly correlated.}

Previously, potential vorticity generation at the shock was studied by \citet{kol03}, who focused on the co-orbital region of the protoplanet, and by \citet{li05}, who numerically investigated the dependence of potential vorticity on spatial resolution in 2D inviscid disks. Both studies addressed the flow instability caused by the potential vorticity generation. Recently, \citet{yu10} explored the time evolution of the potential vorticity and its dependence on $\rs$. \citet{mut10} also studied the possibility of using potential vorticity to identify the shock formation in an attempt to verify the theoretical $\lsh-\mplanet$ scaling relation in GR01. However, their potential vorticity profiles were rather noisy preventing meaningful quantitative comparison.

\citet{lin10} investigated the potential vorticity generation by a massive planet ($\mplanet\gtrsim\Mth$, so the wave shocks immediately after being excited). They followed fluid elements on horseshoe orbits, and confirmed that the potential vorticity is generated as material passes through the two spiral shocks. In a global cylindrical geometry employed in their work, a fluid element gets a kick in the potential vorticity every time it passes a spiral shock, so the potential vorticity in the simulation box increases with time. This is also the case in our 2D local shearing sheet geometry when we switch the $y$ boundary condition to periodic. However, while in \citet{lin10} the potential vorticity stops increasing and reaches a plateau after $30-50$ orbits, in our low mass planet ($\mplanet\ll\Mth$) and high resolution cases we do not see this saturation. In one experiment with 100 orbits, our $\Delta\zeta$ linearly increases with time throughout the entire simulation time. It is not clear what causes this difference, but we suspect that the low resolution (which introduces larger numerical viscosity) might be responsible for it in the \citet{lin10} case.

%%%%%%%%%%%%%%%%%%%%%%%%%%%%%%%%%%%%%%%%%

\subsection{Effect of Equation of State}\label{sec:eos}

\begin{figure}[tb]
%\vspace*{-0.5cm}
\begin{center}
\epsscale{0.45} \plotone{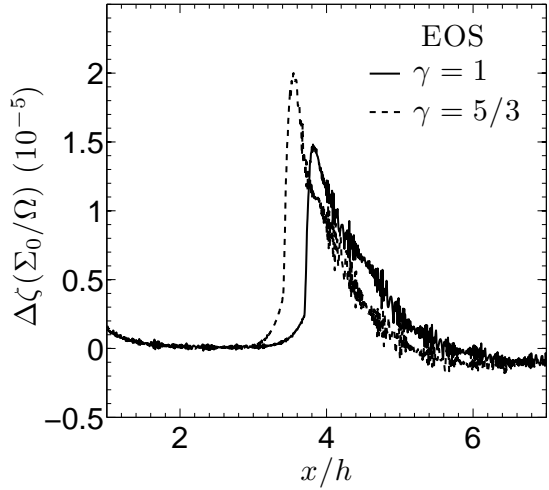}
%\vspace*{-0.5cm}
\end{center}
\figcaption{Radial $\Delta\zeta$ profile for two simulations with otherwise identical parameters (including the adiabatic sound speed) but different EOS ($\gamma=1$ and $\gamma=5/3$). $\mplanet=2.09\times10^{-2}\Mth$ in both cases, which corresponds to $\lsh\approx4h$ for $\gamma=1$ and $\lsh\approx3.6h$ for $\gamma=5/3$.
\label{fig:eos}}
\end{figure}

In linear regime, the density wave evolution does not depends on the equation of state, as long as the sound speed of the gas is fixed. However, the EOS has a prominent effect in the nonlinear regime, which results in a dependence of the shock location on $\gamma$ (Eq. \ref{eq:ls-mp}). We show in Figure~\ref{fig:eos} the radial $\Delta\zeta$ profiles for two simulations with otherwise identical parameters (including the sound speed) but with different EOS ($\gamma=1$ and $\gamma=5/3$). There are two major differences between them. First of all, $\lsh$ for the two cases are different, with larger $\gamma$ resulting in earlier shock. The difference between $\lsh$ in two cases is consistent with the theoretical prediction for these $\gamma$ ($\sim10\%$). Second, the peak $\Delta\zeta$ value in the $\gamma=5/3$ case is about $25\%$ higher than in the $\gamma=1$ case. On the other hand, the decay of the $\Delta\zeta$ profiles is qualitatively similar for different EOS.

%%%%%%%%%%%%%%%%%%%%%%%%%%%%%%%%%%%%%%%%%

\subsection{The $\lsh-\mplanet$ Relation}\label{sec:ls-mp}

\begin{figure}[tb]
%\vspace*{-0.5cm}
\begin{center}
\epsscale{0.45} \plotone{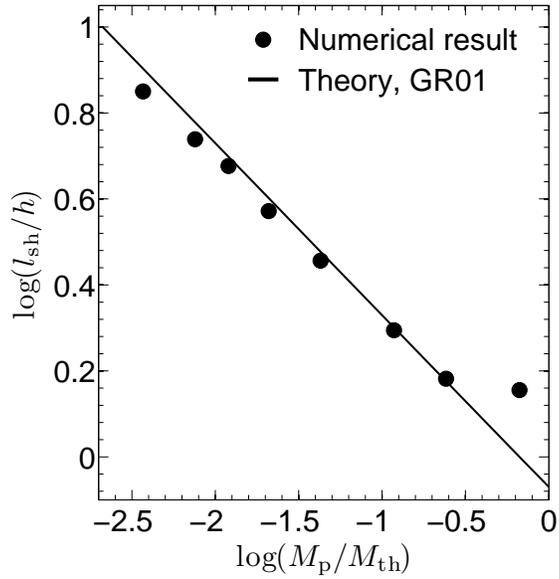}
%\vspace*{-0.5cm}
\end{center}
\figcaption{Numerical $\lsh$ as a function of $\mplanet$ (dots) as well as the theoretical prediction of GR01 (Eq. (\ref{eq:ls-mp}), solid line). $\lsh$ is determined as the midpoint of the $\Delta\zeta$ jump at the shock, as we discussed in \S~\ref{sec:pvnumerical}. For the five low mass cases we use resolution=$256/h$ and $\rs=h/32$, and for the three high mass cases we use resolution=$128/h$ and $\rs=h/16$. The smallest and largest $\mplanet$ here correspond to a few Lunar and Earth mass at 1 AU.
\label{fig:ls-mp}}
\end{figure}

One of the most important results of the analytical theory by GR01 is the relation (\ref{eq:ls-mp}) between the shocking length $\lsh$ and the mass of the planet $\mplanet$. This relation plays a central role in the calculation of every process driven by the nonlinear evolution, such as the migration feedback and gap opening \citep{raf02b}. Here we provide the numerical confirmation of this relation, as shown in Figure~\ref{fig:ls-mp}.

We define the shock location as the value of $x$ at which $\zeta$ reaches half of its maximum value at the jump. The numerical data points nicely agree with the theoretical expectation $\lsh\propto\mplanet^{-2/5}$ for about {\it 2.5 orders of magnitude} in $\mplanet$, with deviation $<10\%$ for most of the $\mplanet$ range. The smallest $\mplanet$ presented here ($3.69\times10^{-3}\Mth$) corresponds to $\sim4$ Lunar mass and the largest $\mplanet$ ($0.667\Mth$) corresponds to $\sim8M_\oplus$ at 1 AU for a MMSN model. The numerical result deviates from theory at the largest $\mplanet$ as expected, because the wave excitation and the shock formation regions are no longer separated there. At the small $\mplanet$ end we also see a trend of deviation from theory. This is because the nonlinearity for such low-mass planets is so small that the linear dissipation due to numerical viscosity becomes non-negligible and starts to damp the wave prior to the theoretically expected location (see discussion in \S~\ref{sec:viscosity}). We expect that modeling the disk at even higher resolution than we have ($256/h$) or more accurate algorithms will resolve this problem (\S~\ref{sec:variety}).

Previous attempts to verify $\lsh-\mplanet$ relation (\ref{eq:ls-mp}) were pioneered by \citet{paar}, who inferred $\lsh$ from the width of the gap opened by the planet. They found $\lsh$ to {\it increase} with decreasing $\mplanet$, in agreement with GR01 theory. However, the resolution of his simulations was insufficient for quantitative verification of Eq. (\ref{eq:ls-mp}). Subsequently, \citet{yu10} confirmed $\lsh-\mplanet$ relation using potential vorticity as means of shock detection. Their calculations spanned only about an order of magnitude in $\mplanet$ and were not fully converged ({\it e.g.} in terms of softening length $\rs$). Since their simulations were run in cylindrical geometry, \citet{yu10} found $\lsh$ in the inner disk to be smaller than in the outer disk. This is to be expected, since global shear rate is higher in the inner disk, which causes faster nonlinear evolution there in agreement with \citet{raf02a}. Nevertheless, the results of \citet{yu10} for $\lsh-\mplanet$ relation are in reasonable quantitative agreement with Eq. (\ref{eq:ls-mp}) of this paper.

%%%%%%%%%%%%%%%%%%%%%%%%%%%%%%%%%%%%%%%%%

\subsection{The decay of Angular Momentum Flux}\label{sec:amfdecay}

%The patten of the decay of the angular momentum flux carried by the density wave is very important in the sense that it determines how the disk will respond to the presence of the protoplanets, which subsequently determines the efficiency of the migration feedback and the gap opening process.

The spatial patten of the decay of the angular momentum flux carried by the density wave determines how the disk will respond to the presence of the protoplanets, which subsequently determines the efficiency of the density feedback and the gap opening. GR01 studied the post-shock AMF decay, and predicted the asymptotic behavior of AMF at large distance (Eq. (\ref{eq:amf-decay}), also see Figure 3 in GR01.). Numerically, we calculate the AMF as:
\begin{equation}
F_H(x) = \Sigma_0\int\limits_{-\infty}^\infty uvdy,
\label{eq:amf-numerical}
\end{equation}
where $u$ and $v$ are the azimuthal and radial velocity perturbation of the fluid with respect to the background shear profile. In theoretical calculation, GR01 showed that $F_H(x)$ is given by
\begin{equation}
F_H(x) = \frac{27\, \cs^3\Sigma_0}{2^{3/2}(\gamma+1)^2\Omega}\left(\frac{\mplanet}{\Mth}\right)^2
~\Phi({t}),\nonumber\\[10pt]
\label{eq:action}
\end{equation}
where the dimensionless AMF $\Phi(t)$ is defined as
\begin{equation}
\quad\Phi({t}) \equiv \int \chi^2(\eta,{t})\,d\eta.
\end{equation}
and $\chi$, $\eta$, and $t$ are defined in Eqs. (\ref{eq:t}-\ref{eq:chi}). Here we follow their notation and measure $\Phi(t)$ from our simulations by calculating $F_H(x)$ and using Eq. (\ref{eq:action}).

\begin{figure}[tb]
%\vspace*{-0.5cm}
\begin{center}
\epsscale{0.45} \plotone{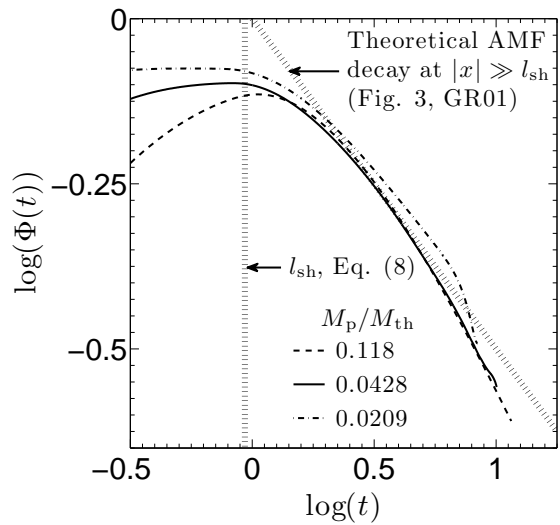}
%\vspace*{-0.5cm}
\end{center}
\figcaption{Numerical result for the AMF decay after the shock formation for three $\mplanet$. The two axes are scaled to facilitate direct comparison with Fig. 3 in GR01 (see \S \ref{sec:amfdecay} for details). For the two low mass cases we use resolution $256/h$ and $\rs=h/32$, and for the highest mass case we use resolution $128/h$ and $\rs=h/16$. Theoretical asymptotic AMF decay scaling relation (Eq. (\ref{eq:amf-decay})) at $x\gg\lsh$ from GR01 and the position of $x=\lsh$ are indicated by dotted lines.
\label{fig:amf-decay}}
\end{figure}

We plot the numerically measured $\Phi({t})$ in Figure~\ref{fig:amf-decay} for three different values of $\mplanet$. Our numerical results agree well with the theoretical prediction at large distance after the shock, which is the working range of the theory, showing that the AMF decay is indeed close to $|x|^{-5/4}$. To the left of the theoretical shocking distance is the linear evolution regime, where AMF increases as the wave accumulates angular momentum. Note that in the linear region, a pattern of AMF increase with $x$ should not depend on $\mplanet$. The apparent dependence is because $t$ in Figure~\ref{fig:amf-decay} has been scaled by $\mplanet^{-1}$. At $\lsh$ the wave AMF stops increasing and starts to dissipate. To the right of $\lsh$, in the nonlinear regime the patten of AMF decay with $x$ does depend on $\mplanet$. However, we remove this dependence and make the AMF decay pattern independent of $\mplanet$ by using $t$ instead of $x$ in Figure \ref{fig:amf-decay}.

%%%%%%%%%%%%%%%%%%%%%%%%%%%%%%%%%%%%%%%%%

\section{Effect of Various Numerical Algorithms}\label{sec:variety}

%%%%%%%%%%%%%%%%%%%%%%%%%%%%%%%%%%%%%%%%%

\begin{figure}[tb]
%\vspace*{-0.5cm}
\begin{center}
\epsscale{0.45} \plotone{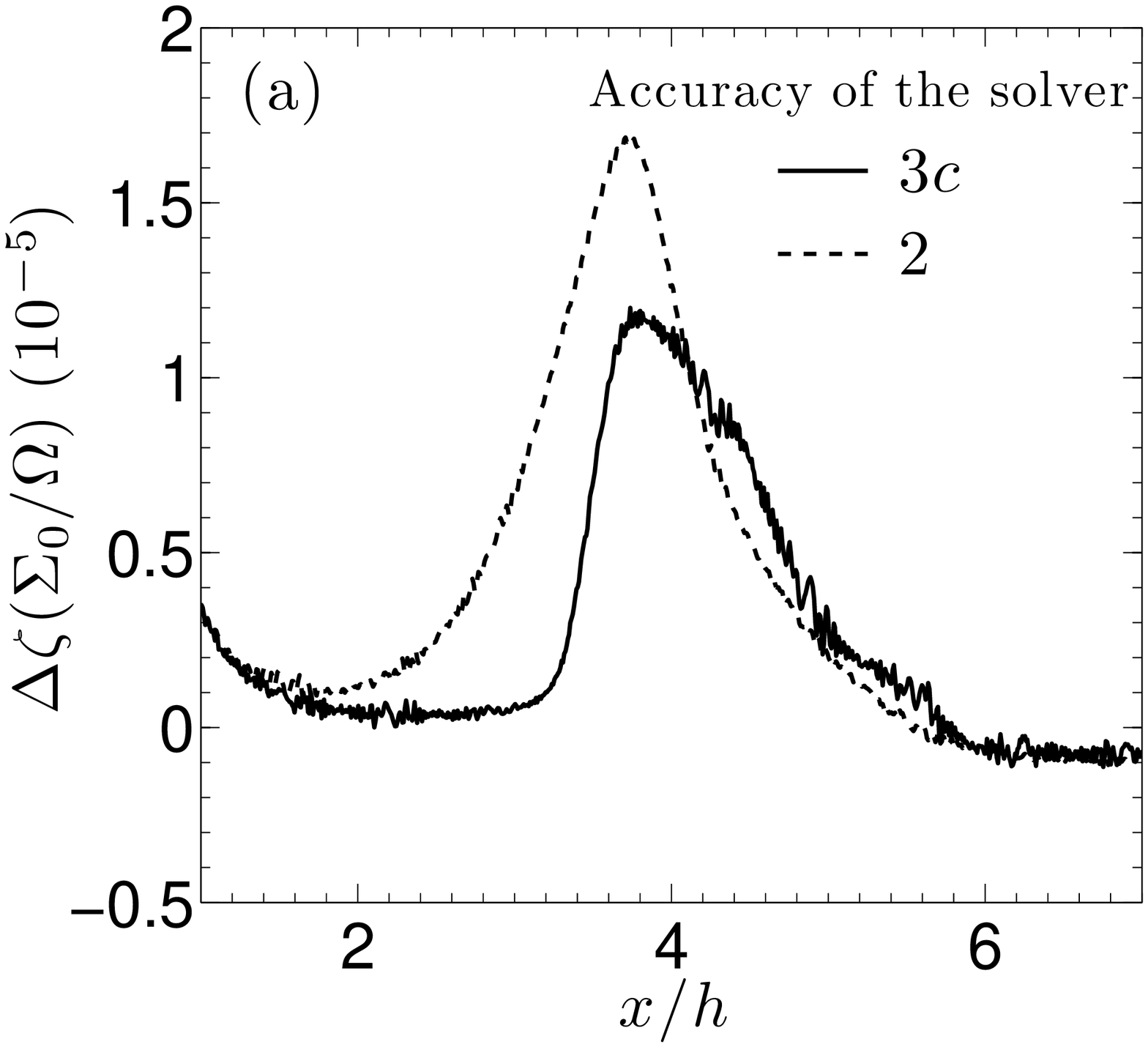} \plotone{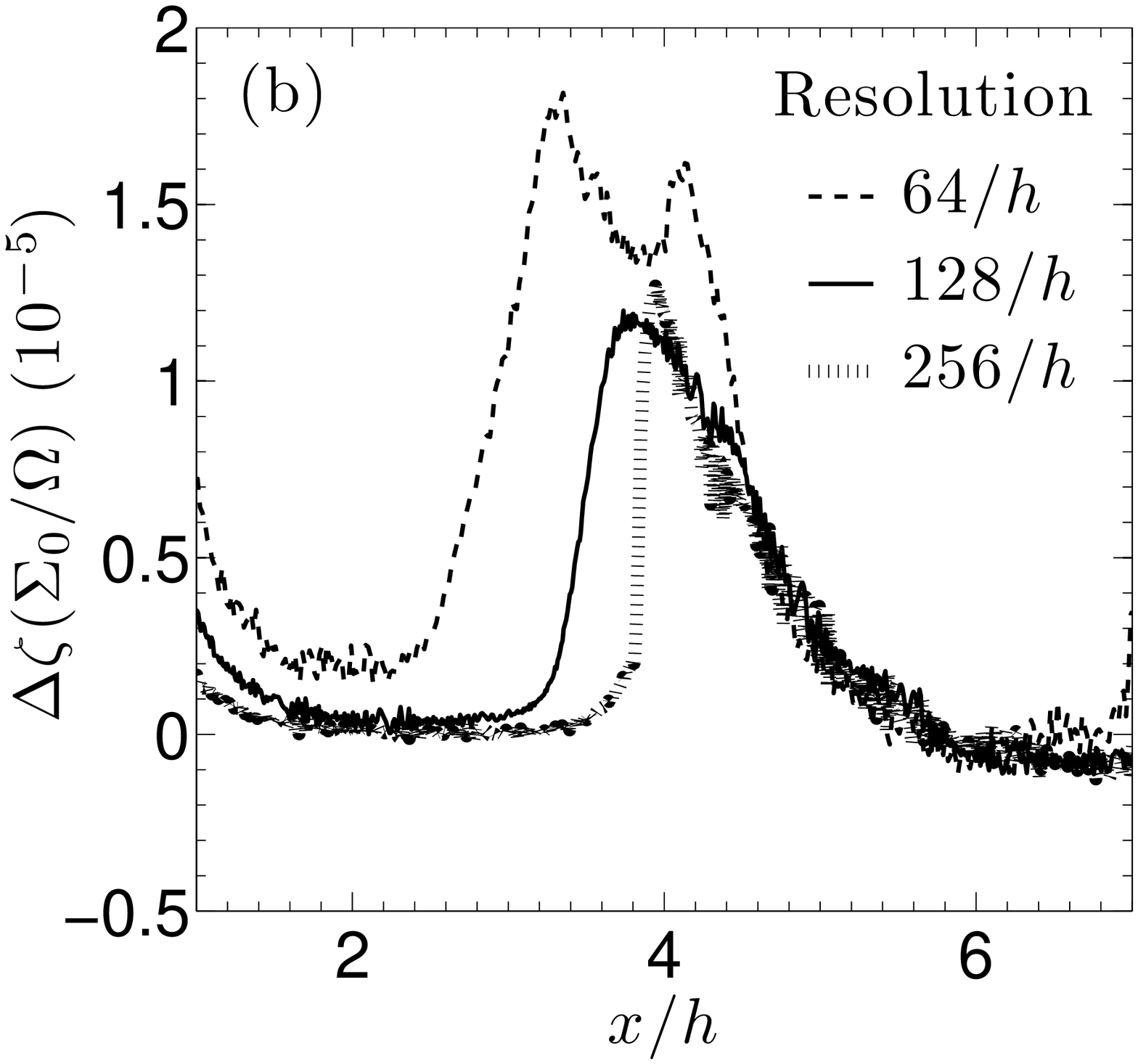} \plotone{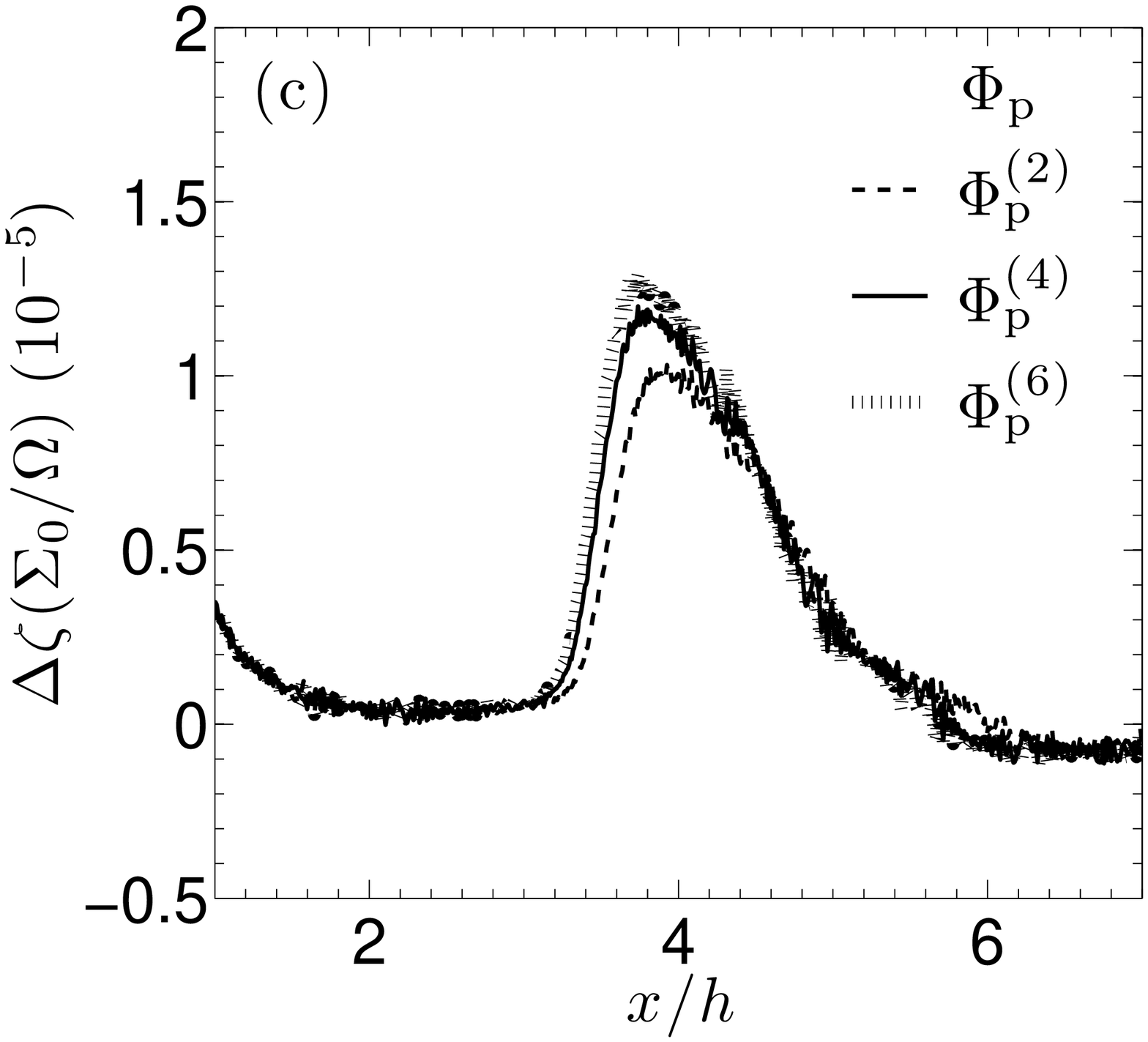} \plotone{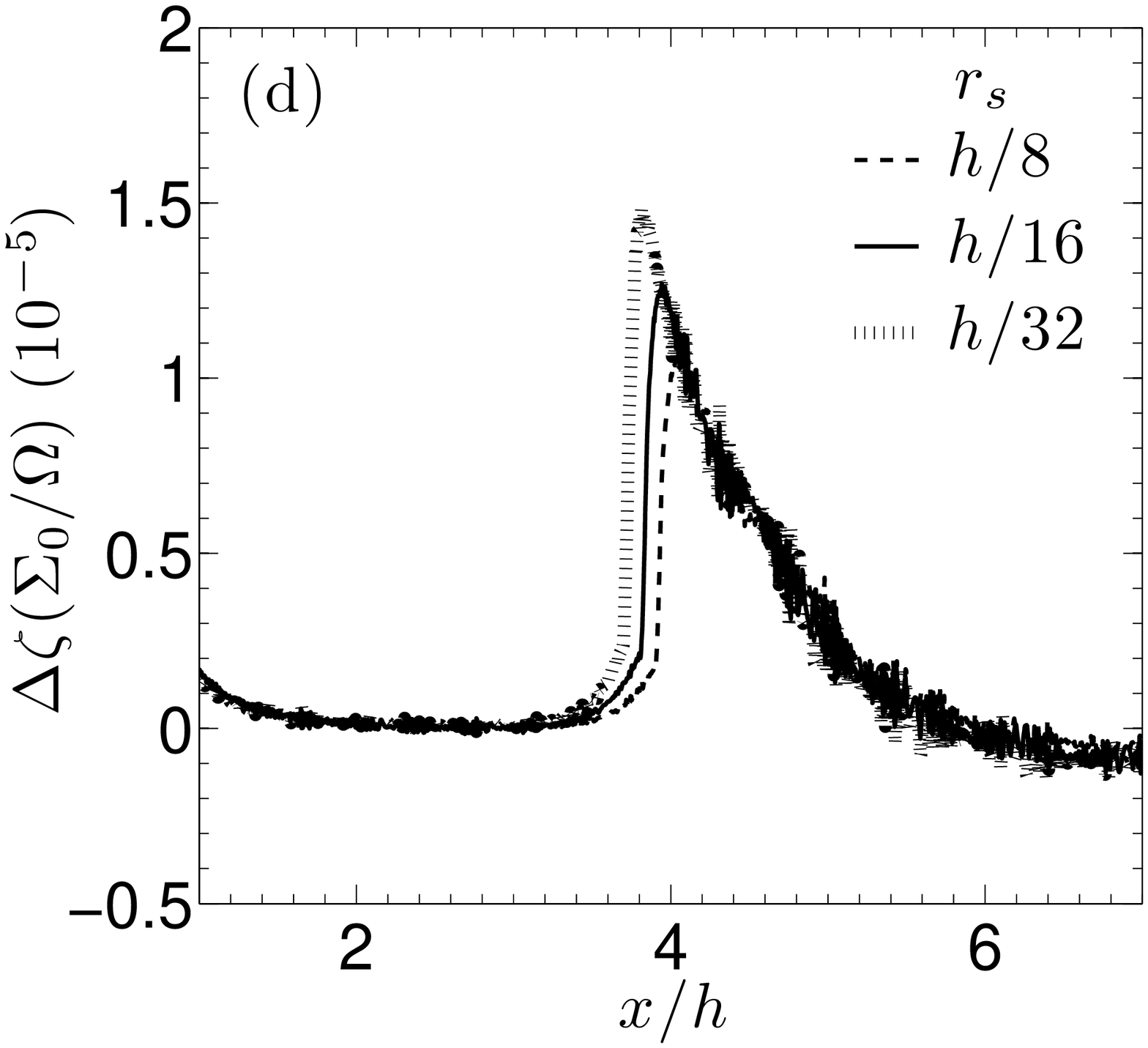}
%\vspace*{-0.5cm}
\end{center}
\figcaption{Radial profile of $\Delta\zeta$ for simulations with different order of accuracy (a, with resolution=$128/h$, $\rs=h/16$), resolution (b, $\rs=h/16$),  $\phip$ (c, with resolution=$128/h$, and $\rs=h/16$), and $\rs$ (d). Other numerical algorithms which are not mentioned are drawn from our fiducial choices (\S \ref{sec:code}). For all simulations we use $\mplanet=2.09\times10^{-2}\Mth$ (corresponding to $\lsh\approx4h$).
\label{fig:pv-variety}}
\end{figure}

In this section, we investigate the impact of various numerical algorithms on the nonlinear wave evolution and the subsequent shock formation by looking at their effect on the generation of potential vorticity. The algorithms we explore are the same as in paper I, as shown in Table~\ref{tab:parameter}, and values corresponding to our standard case are indicated in boldface. Figure~\ref{fig:pv-variety} shows the radial $\Delta\zeta$ profile for different values of various numerical algorithms, and we discuss the effect of each of them in detail below.

%%%%%%%%%%%%%%%%%%%%%%%%%%%%% tab1 %%%%%%%%%%%%%%%%%%%%%%%%%%%%%%%%%%%%%%%%%
\begin{deluxetable}{lc}
\tabletypesize{\scriptsize}
\tablewidth{0pt}
\tablecaption{Algorithms of the simulations}
\tablehead{
Parameters & Range
}
\startdata
Riemann solver (flux function) used & {\bf Roe}, HLLC \\
Order of accuracy & {\tt 2}, {\tt {\bf3}{\bf c}}, {\tt 3p} \\
Boundary conditions in $y$ & Outflow, {\bf Inflow/Outflow} \\
Resolution of the simulation (cells per $h$) & 64, 128, {\bf 256} \\
Planetary potential (see \S \ref{sec:code}) & $\phip^{(2)}$, $\bf \phip^{\bf(4)}$, $\phip^{(6)}$ \\
Softening length & 1/8, 1/16, {\bf 1/32} \\
Equation of states of the fluid & {$\bf \gamma=1$}, $\gamma=5/3$ \\
Mass of the planet (0.01$\Mth$) & 66.7, 24.2, 11.8, 4.28, 2.09, 1.20, 0.757, 0.369 \\
\enddata
\label{tab:parameter}
\end{deluxetable}
%%%%%%%%%%%%%%%%%%%%%%%%%%%%% end tab1 %%%%%%%%%%%%%%%%%%%%%%%%%%%%%%%%%%

%%%%%%%%%%%%%%%%%%%%%%%%%%%%%%%%%%%%%%%%%

\subsection{Solver and its accuracy}\label{sec:solver}

In our simulations we compare two different Riemann solvers --- Roe's linearized solver (\citealt{roe81}) and HLLC \citep{tor99} --- with three different algorithms for the spatial reconstruction step \citep{sto08}: second order with limiting in the characteristic variables (denoted {\tt 2} in this work), which is the predominant choice in literature in this field, and third order with limiting in either the characteristic variables ({\tt 3c}), or in the primitive variables ({\tt 3p}).

Similar to their effect in the linear regime (paper I), the two kinds of solvers yield almost identical results on the $\Delta\zeta$ profile. On the other hand, different orders of accuracy do make significant differences. {\tt 3p} accuracy (not shown in this figure) introduces large high frequency fluctuations on $\Delta\zeta$ profile compared to {\tt 2} and {\tt 3c} accuracy. For the two spatial reconstruction steps with limiting in the characteristic variables, as shown in Figure~\ref{fig:pv-variety}a {\tt 2} accuracy generates a much smoother rise of the $\Delta\zeta$ profile around $\lsh$, compared with the sharp jump in the {\tt 3c} accuracy case. In addition, {\tt 2} case advances shock formation, causing the numerical $\lsh$ to disagree with theoretical prediction (\ref{eq:ls-mp}). Analogous to Paper I, we find that the effect of reducing the accuracy from third order to second order is very similar to reducing the resolution by a factor of 2.

A critical ingredient to any total variation diminishing (TVD) reconstruction scheme, as are used in Athena, are the slope limiters used to ensure monotonicity. We have not explored the use of different limiter in this work, but this may affect the accuracy of the {\tt 3p} and {\tt 3c} methods.

%%%%%%%%%%%%%%%%%%%%%%%%%%%%%%%%%%%%%%%%%

\subsection{Resolution}\label{sec:resolution}

Resolution has a strong influence on the shock formation, as shown in Figure~\ref{fig:pv-variety}b. Lower resolution considerably accelerates shock formation, causing $\lsh$ to deviate from the theoretical prediction. Furthermore, the shape of the rising edge of the $\Delta\zeta$ profile depends on resolution. Increasing resolution steepens the edge of the $\Delta\zeta$ curve, and eventually makes it almost vertical in the case of $256/h$ (which is a sign that the convergence on resolution has been achieved). We also find that at low resolution ($64/h$ in our case), the $\Delta\zeta$ profile always demonstrates a double peaked structure, which is a purely numerical artifact.
% While the left peak is the persistent one, the second peak at larger $x$ only manifests itself when resolution is below some critical value ($\lesssim64/h$ in our simulations), though the two are close to each other. This is a numerical issue which may be related to the fact that the low resolution simulations can not resolve the motion of the fluid around the planet.

%%%%%%%%%%%%%%%%%%%%%%%%%%%%%%%%%%%%%%%%%

\subsection{Planetary potential}\label{sec:potential}

Following paper I, we test the sensitivity of our results to the specific method of potential softening, and try three different forms of the softened planetary potential. One of them is the second order potential defined as
\begin{equation}
\phip^{(2)}=-\g \mplanet\frac{1}{(\rho^2+\rs^2)^{1/2}},
\label{eq:phi2}
\end{equation}
which converges to $\Phi_K=-\g\mplanet/\rho$ at $\rho\gg \rs$ as $\left(\rs/\rho\right)^2$ (which means the fractional error is O$((r_s/\rho)^2)$ as $r_s/\rho\rightarrow0$). This is the potential  most commonly used in numerical hydrodynamic studies. We also studied the fourth order potential (Eq. \ref{eq:phi4}), converging to the point mass potential as $\left(\rs/\rho\right)^4$ for $\rho\gg \rs$, and the sixth order potential:
\begin{equation}
\phip^{(6)}=-\g \mplanet\frac{\rho^4+2.5\rho^2\rs^2+1.875\rs^4}{(\rho^2+\rs^2)^{5/2}}
\label{eq:phi6}
\end{equation} 
converging to $\Phi_K=G\mplanet/\rho$ as $\left(\rs/\rho\right)^6$ at $\rho\gg \rs$. We also varied softening length $\rs$ at fixed form of the potential. The effect of various $\phip$ is shown in panel (c), and that of different $\rs$ in panel (d). We find that switching to $\phip$ one level higher in accuracy increases $\Delta\zp$ by $\sim10-15\%$, steepens the edge of the $\Delta\zeta$ profile, and slightly advances the shock formation. Reducing $\rs$ by a factor of 2 has quantitatively similar effects. But in general the variation of $\phip$ and $\rs$ has far less prominent effect on the $\Delta\zeta$ evolution than the accuracy of solver and resolution.

Based on this discussion we can state that in order to accurately follow the nonlinear wave evolution and capture the shock formation, it is crucial to use high order of accuracy of the numerical solver, and high spatial resolution. Simulations which do not satisfy these criteria may be affected by numerical artifacts and may not be able to resolve properly either the nonlinear wave evolution or its consequences --- the migration feedback and the gap opening.

%%%%%%%%%%%%%%%%%%%%%%%%%%%%%%%%%%%%%%%%%

\section{Nonlinear Evolution in Presence of Explicit Viscosity}\label{sec:viscosity}

Nonlinear evolution of density waves launched by the low-mass planets can be significantly affected by linear damping even if the latter is due to numerical viscosity. This is because the time it takes for nonlinearity to have an effect is longer for smaller $\mplanet$, which enhances the relative contribution of the linear wave damping.

\begin{figure}[tb]
%\vspace*{-0.5cm}
\begin{center}
\epsscale{0.41} \plotone{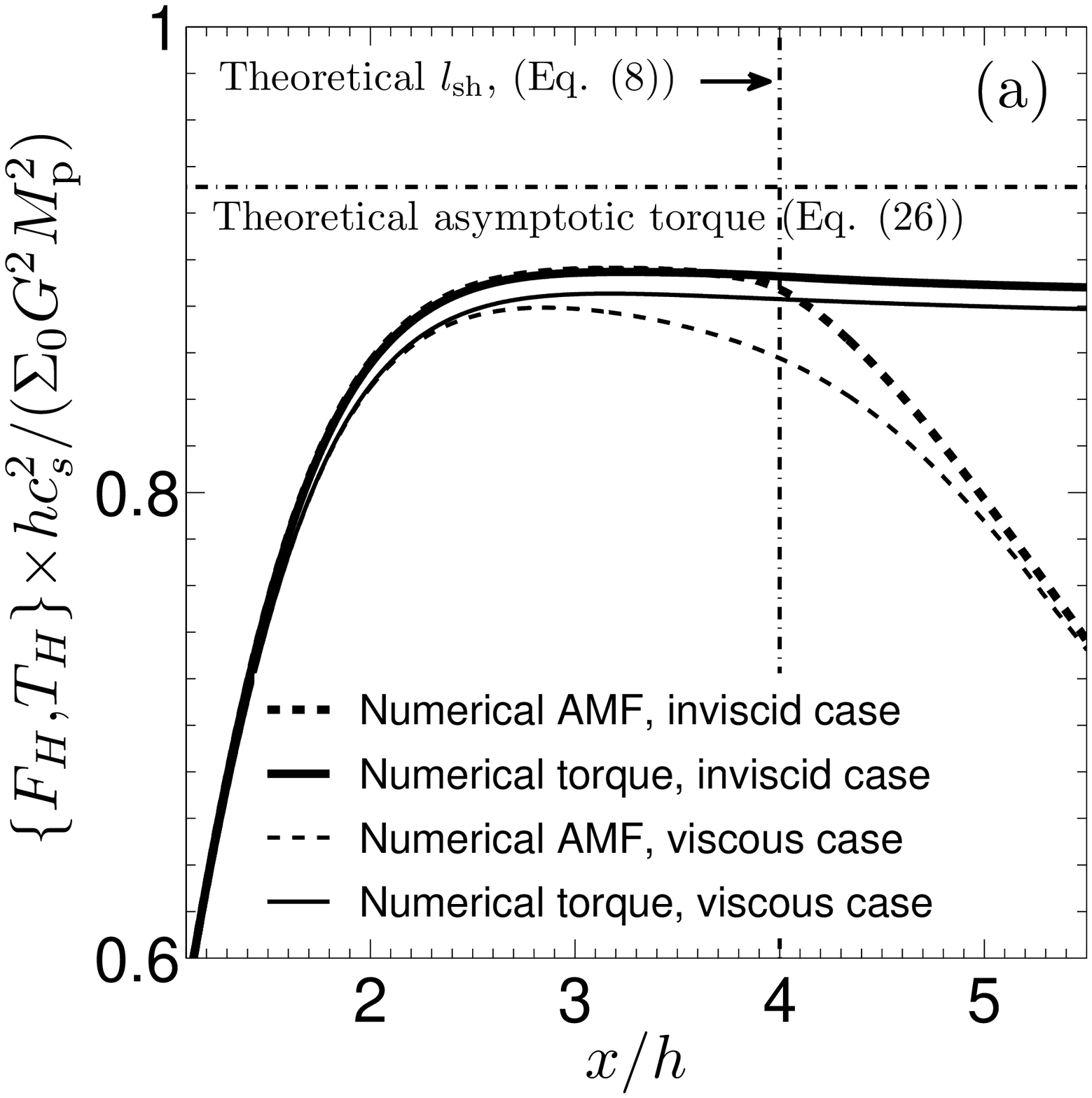}\epsscale{0.45}  \plotone{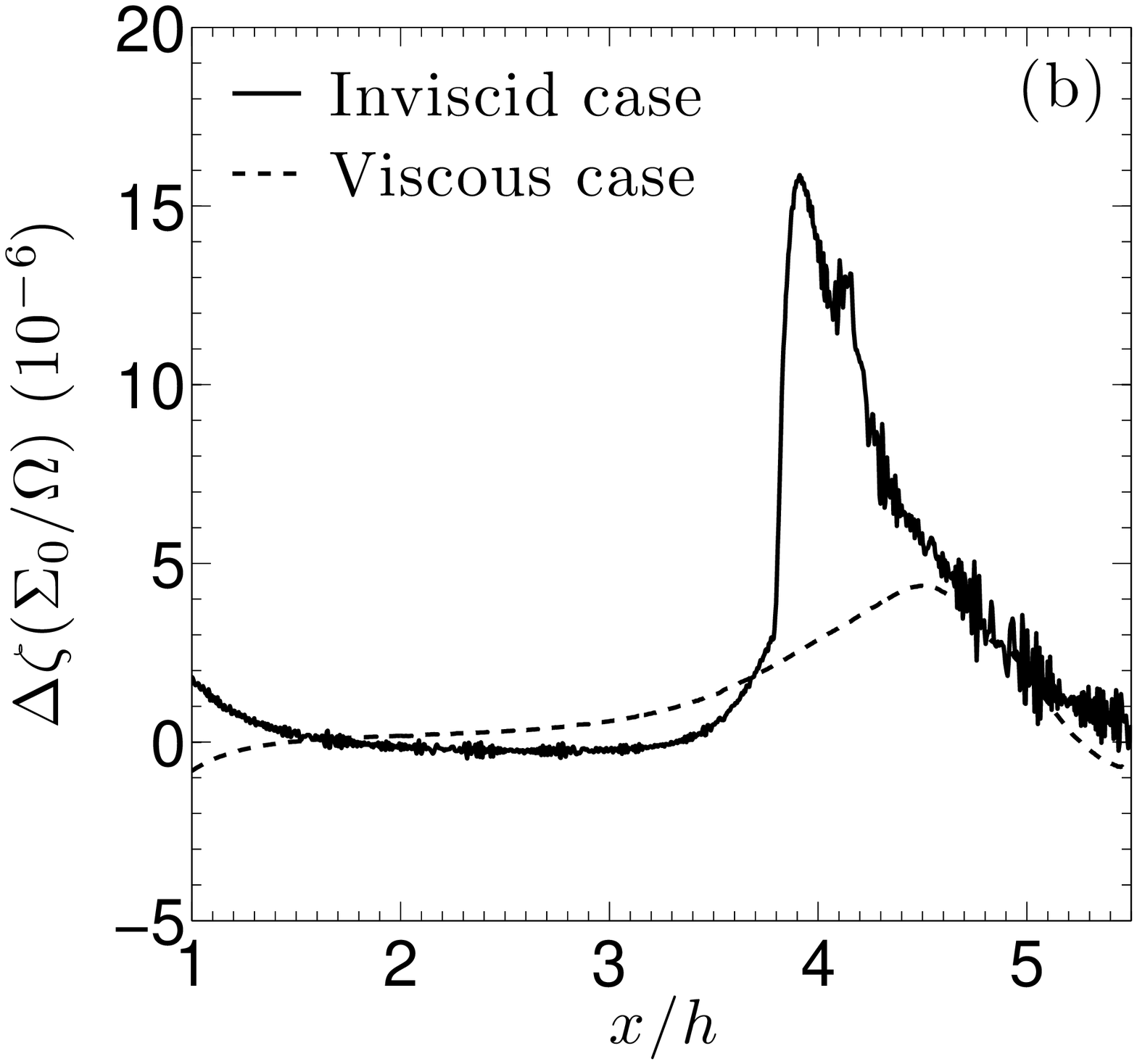}
%\vspace*{-0.5cm}
\end{center}
\figcaption{The effect of explicit viscosity in our runs. The viscous simulation is done with an explicit Navier-Stokes viscosity (Shakura-Sunyaev $\alpha=\nu/(h\cs)=10^{-4}$), and the inviscid run is our standard simulation without explicit viscosity. Both simulations are done with $\mplanet=2.09\times10^{-2}\Mth$ (corresponding to $\lsh\approx4h$). Panel (a) shows the numerical AMF and torque calculation, and panel (b) shows the radial $\Delta\zeta$ profile.
\label{fig:viscosity}}
\end{figure}

To explore the effect of linear damping on nonlinear wave evolution, we carry out an experiment with explicit Navier-Stokes viscosity. We choose kinematic viscosity $\nu$ to correspond to Reynolds number $Re\equiv h\cs/\nu=10^4$. This is equivalent to having effective Shakura-Sunyaev $\alpha=\nu/(h\cs)=10^{-4}$. We compare the results with another simulation under otherwise identical conditions but without explicit viscosity (our numerical viscosity is at least 10 times smaller in this case, which corresponds to Shakura-Sunyaev $\alpha\lesssim10^{-5}$). The results of the comparison are shown in Figure~\ref{fig:viscosity}.

%\rdnote{in paper I we did present a similar figure to show the effect of low resolution. The reason I still show panel a here is, (1) the effect of numerical viscosity and explicit viscosity is not identical, although the general trend is similar, and (2), more importantly, I have a feeling that most people think $\alpha=10^{-4}$ is a very small viscosity,, which $\alpha=10^{-4}\sim0$, here we can show that the effect of $\alpha=10^{-4}$ on AMF and torque is actually quite strong. While low resolution $\sim$ high viscosity, it's hard to quantify its effect and assign an $\alpha$ to a low resolution. So I suggest we still keep the discussion here.}

As discussed in Paper I, before the shock the numerical AMF and accumulated torque calculations should agree with each other, because of the angular momentum conservation. After the shock formation, dissipation damps the wave and transfers the angular momentum from the wave to the local disk material causing the AMF to drop below the accumulated torque curve. Accumulated numerical torque in our simulations is calculated as:
\begin{equation}
T_H(x) = \int\limits_0^x\frac{dT_H}{dx}dx =  \int\limits_0^xdx\int\limits_{-\infty}^\infty dy\delta\Sigma\frac{\partial\phip}{\partial y},
\label{eq:specific torque}
\end{equation}
and its asymptotic value in the linear case is (GT80):
\begin{equation}
T_{H,\rm final} \approx 0.93\left(G \mplanet\right)^2\frac{\Sigma_0\Omega_p}{c_s^3}.
\label{eq:T_H_tot}
\end{equation}

As shown in the panel (a), while in the inviscid case this deviation happens right at the theoretically predicted shocking length ($\lsh\approx4h$), in the viscous case AMF starts to deviate from the $T_H(x)$ much earlier (at $\sim2.5h$). This indicates that dissipation happens earlier in the viscous case due to the linear damping, which then must be reflected in the disk density redistribution, eventually affecting migration feedback and gap opening. In addition, panel (b) shows that in viscous case $\Delta\zeta$ starts to rise from zero much earlier than in the inviscid case, also indicating premature dissipation. Furthermore, the rise of $\Delta\zeta$ in the viscous case is very gradual, in contrast to the sharp jump in the inviscid case, and the peak $\Delta\zp$ is also significantly reduced (by a factor of $\sim4$). The combination of the two effects makes the shock detection very ambiguous in the viscous case.

We note that normally disks with $\alpha=10^{-4}$ are considered to have very low viscosity, if not inviscid. However, as we show here, the physics of the nonlinear density wave evolution is very subtle so that even an $\alpha=10^{-4}$ viscosity can dramatically affect wave damping. In simulations which aim at investigating the nonlinear wave evolution, the low spatial resolution ($\leq32/h$, typically employed in the literature) will likely introduce high numerical viscosity. This is likely to give rise to inaccurate numerical results in a way similar to our viscous simulation shown here, and via the improperly captured back reaction on the disk density distribution, affect the description of the migration feedback and gap opening.

%%%%%%%%%%%%%%%%%%%%%%%%%%%%%%%%%%%%%%%%%

\section{Implications for realistic protoplanetary disks}
\label{sec:implications}

%%%%%%%%%%%%%%%%%%%%%%%%%%%%%%%%%%%%%%%%%

The analytical and numerical calculations presented in this work directly 
apply only to the case of a 2D, laminar disk, containing a single 
low mass planet. We now discuss the applicability of our results to
more realistic protoplanetary disks.

First, we note that for typical $h/\rp$ values of protostellar disks (on the order of 0.1, \citealp{har98,chiang}), the size of the radial domain in this work (up to $10h$ on one side) would imply global disk dimensions, which might be seen as contradictive to the essence of the shearing box simulation. However, our numerical calculation is an idealization needed for checking the GR01 theory. Extensions of the GR01 theory to the global case are available in \citet{raf02a}, with which global simulations should be compared. Here we just point out that qualitatively things are the same in the global case.

Excitation of density waves by planets in fully three-dimensional (3D)
disks has been previously investigated by a number of authors 
\citep{lubow93,koryc95,takeuchi98}. These studies have generally
shown that disks with vertical thermal stratification do not
support the modes similar to the modes existing in a purely 2D 
disk \citep{koryc95,lubow98,ogi99}. Thus, the results 
of our study cannot be directly applied to such thermally 
stratified disks.

However, protoplanetary disks at advanced stages of planet formation
are expected to be passive \citep{chiang}, heated predominantly 
by the radiation of their central stars with only negligible 
contribution from accretional energy release. Because of the external 
illumination such disks are expected to be vertically isothermal. 
It was previously found \citep{lubow93,takeuchi98} that in 
vertically isothermal disks planetary gravity always very  
effectively excites the two-dimensional mode with no vertical motion, 
which is similar to the density wave in a 2D disk studied here. Even 
though other modes with non-zero vertical velocity perturbation 
are also excited by the planet in 3D isothermal disks,
these modes are found to carry only a small fraction
of the total angular momentum flux \citep{takeuchi98}. Thus, 
density wave excitation and nonlinear dissipation in realistic 
passive protoplanetary disks should be very similar to the picture 
outlined in GR01 and this work.

The assumption of the laminar inviscid background flow in the disk 
used in this work may be violated in realistic protoplanetary disk 
if it is strongly turbulent, e.g. due to the operation of the 
magnetorotational instability (MRI). However, this instability 
and associated deviations from the purely laminar flow are expected 
to be greatly suppressed in the so-called ``dead zones'' of
protoplanetary disks \citep{gam96}, which are so weakly ionized that 
the operation of MRI is not supported there because of non-ideal 
MHD effects \citep{fleming}. These zones are expected to extend over
the several-AU wide region where planet formation is expected to occur
\citep{turner} and in this part of the disk our results obtained 
under the assumption of laminar flow should be applicable. One should 
also mention that because of the suppressed viscous dissipation inside
the dead zone the assumption of a vertically isothermal thermal 
structure of these regions is well justified.

Protoplanetary disks may easily host a number of protoplanets
simultaneously and one may wonder whether the overlap of the density
waves excited by different objects can have some effect on their
evolution. This effect is not important in the limit of low
mass planets $M_p\lesssim \Mth$ explored in this work since the
density waves excited by such objects are only {\it weakly}
nonlinear. As a result, the possible overlap of such density waves 
is, to a good approximation, just a linear combination of the 
independent density waves excited by individual planets. To summarize, the results of our work should
directly apply at the very least to the density wave evolution 
in low viscosity regions (``dead zones'') of passive protoplanetary 
disks heated predominantly by their central stars.

%%%%%%%%%%%%%%%%%%%%%%%%%%%%%%%%%%%%%%%%%

\section{Summary}\label{sec:summary}

%%%%%%%%%%%%%%%%%%%%%%%%%%%%%%%%%%%%%%%%%

We present a numerical study of the nonlinear density wave evolution, based on 2D local shearing box hydrodynamical simulations using the grid-based code Athena. The nonlinear wave evolution plays an important role in the disk-planet interaction and planetary migration. It provides an efficient and robust wave damping mechanism, working in regimes where linear dissipation mechanisms fail, and leading to gap opening by very low mass planets ($\sim M_\oplus$). Nonlinear wave evolution leads to shock formation, wave damping, and transfer of the angular momentum carried by the wave to the local disk material. The latter drives global disk evolution by redistributing the mass in the disk, and leads to the migration feedback and gap opening.

We analytically study the potential vorticity generation at the shock location, and verify our calculation numerically. The scaling relation between $\Delta\zp$ and $\mplanet$ derived from the simulations is very close to the analytical result. We use the jump in potential vorticity as a way of pinpointing the shock location, based on which we numerically confirm with high accuracy (better than $10\%$ in most of the $\mplanet$ range) the theoretically predicted $\lsh-\mplanet$ relation for $\mplanet$ varying by 2.5 orders of magnitude, from few lunar masses to several Earth masses at 1 AU. In addition, the theoretical dependence of the shocking length $\lsh-\mplanet$ relation on the equation of state of the gas is also verified.

We investigate the evolution of the density wave profile in the post-shock regime, and observe its evolution towards the N-wave shape. We verify the theoretical prediction for the evolution of the peak amplitude and the azimuthal width of the N-wave. The post-shock decay of the angular momentum flux carried by the wave also agrees with theory.

Furthermore, the effect of various numerical algorithms on the simulation results is explored, including numerical solver and its accuracy, resolution, planetary potential and the softening length. We find that in order to accurately follow the nonlinear wave evolution and capture the shock formation, high order of accuracy for the solver and high resolution are crucial. Simulations which do not satisfy these criteria will have trouble resolving shock formation and post-shock wave evolution driven by nonlinearity. The latter may further affect the consequences of the nonlinear wave evolution, such as the migration feedback and gap opening by low mass planets.

In addition, we find that the linear viscous damping strongly affects shock formation and the nonlinear wave evolution. Using an experiment with explicit viscosity, we find that the viscosity at the level of $\alpha\sim10^{-4}$ can strongly modify the generation of potential vorticity at the shock, and accelerate the dissipation of the angular momentum carried by the wave. In low resolution simulations, the resulting high numerical viscosity may lead to similar effects on the nonlinear wave evolution and negatively impact the results on disk-planet interaction and planetary migration.

\section*{Acknowledgments}

We are grateful to Jeremy Goodman, Takayuki Muto, Mikhail Belyaev, Cristobal Petrovich, and an anonymous referee for useful discussions and comments. The financial support of this work is provided by NSF grant AST-0908269, and a Sloan Fellowship awarded to RRR.

\clearpage

\end{document}